\documentclass[draft]{agujournal2019}
\usepackage{url} 
\usepackage[inline]{trackchanges} 
\usepackage{soul}
\draftfalse

\journalname{JGR: Atmospheres}

\begin{document}

%
%


\title{The Evolution of Turbulence Producing Motions in the ABL Across a Natural Roughness Transition}

%
%




\authors{Justin P. Cooke\affil{1}, Douglas J. Jerolmack\affil{1,2}, \& George Ilhwan Park\affil{1}}

\affiliation{1}{Department of Mechanical Engineering and Applied Mechanics, University of Pennsylvania, Philadelphia, Pennsylvania, USA}
\affiliation{2}{Department of Earth and Environmental Science, University of Pennsylvania, Philadelphia, Pennsylvania, USA}





\correspondingauthor{George Ilhwan Park}{gipark@seas.upenn.edu}




\begin{keypoints}
    \item When atmospheric flows encounter roughness transitions an Internal Boundary Layer forms and changes the structure of the flow within
    \item Internal boundary layers set the new length-scales of the flow and control turbulence producing motions
    \item The structure of turbulence producing events is shown to scale with Internal Boundary Layer height with implications for sediment transport
\end{keypoints}

%
%

%
%


\begin{abstract}
Landforms such as sand dunes act as roughness elements to Atmospheric Boundary Layer (ABL) flows, triggering the development of new scales of turbulent motions. 
These turbulent motions, in turn, energize and kick-up sand particles, influencing sediment transport and ultimately the formation and migration of dunes -- with knock on consequences for dust emission.
While feedbacks between flow and form have been studied at the scale of dunes, research has not examined how the development of an Internal Boundary Layer (IBL) over the entire dune field influences sediment-transporting turbulence.
Here, we deploy large-eddy simulation of an ABL encountering a natural roughness transition: the sand dunes at White Sands National Park, New Mexico. 
We analyze turbulence producing motions and how they change as the IBL grows over the dune field. Frequency spectrum and Reynolds shear stress profiles show that IBL thickness determines the largest scales of turbulence. More, the developing IBL enhances the frequency, magnitude and duration of sweep and ejection events -- turbulence producing motions whose peaks systematically migrate away from the wall as the IBL thickens.
Because sweep and ejection events are known to drive sediment transport, our findings provide a mechanism for coupling the co-evolution of the landscape and the ABL flow over it.
More broadly, our results have implications for how roughness transitions influence the transport of pollutants, particulates, heat, and moisture.
\end{abstract}


%
%

%


%
%
%
%

\section{Introduction}

The Atmospheric Boundary Layer (ABL) is the $O(10^2-10^3)$ m thick region near the Earth's surface that most feels the shear stresses of the flow~\cite{abedi2021numerical,bouzeid2020persistent,gul2022experimental}.
Owing to the large length-scales and the heterogeneous topography of the Earth, 
ABL flows are highly turbulent, with the friction Reynolds numbers $Re_\tau$ $\sim O(10^6-10^7)$. 
Complex terrain, composed of mountains, forests, hills, and valleys, as well as areas composed of human-made structures, such as urban areas and wind farms, act as roughness elements within the ABL, and present challenges to our understanding~\cite{vanderwel2019turbulent,bouzeid2020persistent,li2023mean}.
In cities, public health is impacted by the interactions between the ABL and buildings, influencing fluxes of heat and pollutants~\cite{li2013synergistic,manoli2019magnitude,li2023mean}.
The growing intensity of wildfires in North America has also shown the importance of the ABL in transporting smoke, with cascading consequences for public health~\cite{zhou2023where,gould2024health,ceamanos2023remote}.
The presence of roughness induces changes to the transport of particles, heat, and momentum within the ABL -- especially near the Earth's surface -- so understanding the physical mechanisms of these changes is critical for prediction and modeling capabilities.

Aeolian sand dunes, which can reach roughness heights $(k)$ taller than 100m ~\cite{gunn2022what}, impede the flow of the ABL, and alter coherent flow structures~\cite{wiggs2012turbulent,anderson2014numerical,wang2019turbulence,bristow2021unsteady}.
The spatial variation in flow over dune topography results in spatial patterns in sand transport across dunes \cite{bagnold1941physics}, which are ultimately connected to the structure and intensity of turbulence forced by roughness \cite{bauer2013critical, bristow2021unsteady,tan2023turbulent}.  
Entrainment of sand is often driven by turbulent {low-speed fluid moving away from the wall (ejections)} and, once a threshold velocity is met, sand is transported primarily by {high-speed fluid moving toward the wall (sweeps)} \cite{pope,sterk1998effect,leenders2005wind,wiggs2012turbulent, bauer2013critical, rana2021entrainment}.
Sand entrainment, and the associated turbulent motions, are also responsible for dust emission \cite{kok2012physics, parajuli2016new, zhang2022impact}.

Sand dunes have been shown to increase the frequency and intensity of turbulent events that drive sand transport \cite{bristow2021unsteady,tan2023turbulent}. Due to the challenges and complications associated with collecting time- and height-based velocity data over large areas in field campaigns~\cite{abedi2021numerical,bell2020confronting}, and the stringent costs of simulating large-scale atmospheric flows~\cite{bouzeid2020persistent,anderson2020large}, much of the prior work on flows over dunes has been restricted to isolated~\cite{wiggs2012turbulent,bristow2021unsteady,bristow2022topographic} or small clusters~\cite{anderson2014numerical,wang2019turbulence} of dunes. 
What happens, however, when flow over a smooth surface encounters a train of dunes?
Step changes in surface roughness (smooth-to-rough, rough-to-smooth, rough-to-rougher, and vice versa) trigger the formation an Internal Boundary Layer (IBL) \cite{antonia1971response,antonia1972response,hanson2016development,gul2022experimental}.
Within the IBL, the flow gradually adapts to the new near-surface condition, while outside, the flow largely retains the characteristics of the upstream boundary layer. 
{Prior work from }\citeA{gul2022experimental} showed that the IBL acts as a `shield', as the outer region of their rough-to-smooth flow loses spectral energy. 
Due to the increased roughness, we expect that sand dunes will extract more momentum from the flow compared to a smooth surface; and therefore that the wind speed and associated turbulent stresses will progressively decline as the IBL thickens \cite{jerolmack2012internal, gunn2020macro}. 
How does this manifest, in terms of the spatial pattern of turbulence producing motions, over an entire dune field?

White Sands National Park (New Mexico, USA) presents a case study in how development of an IBL, triggered by a roughness transition, can lead to feedbacks between flow and form that affect the evolution of an entire dune field \cite{jerolmack2012internal}. 
In this landscape a quasi-unidirectional wind encounters an abrupt roughness increase from smooth playa to sand dunes, whose size grows rapidly over the first several kilometers and then gradually declines over the next 10 km. 
The observed spatial pattern of sand flux is broadly similar; first increasing rapidly and then gradually declining moving downwind \cite{jerolmack2012internal, gunn2020macro}. 
Spatially sparse wind observations are consistent with a simplified analytical model, that describes a pattern of gradual downwind decrease in surface-wind velocity due to a developing IBL \cite{gunn2020macro}. 

Our prior work used {Wall-Modeled }Large Eddy Simulations ({WM}LES) over White Sands topography in order to resolve the complex flows that give rise to IBL development across the dune field \cite{cooke2024mesoscale}. 
{Using these simulations, we calculated the IBL height for a spatially heterogeneous smooth to rough roughness transition.}
These simulations substantiated the importance of IBL development on sand-dune dynamics, while revealing a novel self-similar structure for the vertical Reynolds stress profile that is scaled by the height of the IBL. 
{More, our work used an amplitude modulation analysis} \cite{mathis2009large} {to quantify -- with the Amplitude Modulation Coefficient $R_{AM}$ -- the influence of the scale-interaction between the large- and small-scale motions.}
{We present these previous results to frame the study of interest: how do turbulence producing motions evolve downstream of the roughness transition, and why is $\delta_i$ the critical length-scale for these flows?}

Here we {run additional simulations to capture new data from the same LES configuration as our previous study}~\cite{cooke2024mesoscale} to investigate the evolution of turbulence producing motions that arise in the developing IBL across the White Sands dune field. We first provide a brief overview of the theory behind the analysis. 
We then describe the wall-modeled LES deployed to simulate a neutrally buoyant ABL flow encountering a spatially heterogeneous roughness transition, including a domain sensitivity analysis. 
We find that both the magnitude and height (above the bed) of turbulence producing motions are set by the scale of the developing IBL, which creates a spatial pattern of turbulence across the dune field. We consider the implications of this pattern for sediment transport and dune dynamics at White Sands, and also more broadly for other landscapes. 

\section{Theory}
\label{sec:theory}

\subsection{Internal Boundary Layer Height Estimation}
\label{ssec:iblHeightEst}

For flows encountering a roughness transition, there are simple correlations for predicting the evolution of the IBL height ($\delta_i$) as a function of the upstream and downstream roughness parameters, $z_{01}$ and $z_{02}$, respectively~\cite{elliott1958growth,townsend1965response,panofsky1973tower,wood1982internal,panofsky1984atmospheric,pendergrass1984dispersion,savelyev2001notes}. 
These correlations were recently tested in detail by~\citeA{gul2022experimental}, and the simplest reasonable model is of the form

\begin{equation}
    \frac{\delta_i}{z_{0}} = a\bigg(\frac{\hat{x}}{z_{0}}\bigg)^{b}.
    \label{eqn:ibl_corr}
\end{equation}

\noindent Here, $a$ and $b$ are empirical constants, and $\hat{x} = x - x_0$ is the streamwise distance from the location $x_0$ where the roughness transition occurs.
The roughness parameter, $z_0$, may be chosen to be the larger of the two, or simply the downstream value. 
Typically, $b = 0.8$, but a wide range of values ($b = 0.2-0.8$) have been reported in the literature~\cite{gul2022experimental}
Measured data of developing IBL thickness $\delta_i$ are often fit to an equation of the form in Equation~\ref{eqn:ibl_corr}.

What defines the boundary of the IBL? Many methods to determine $\delta_i$ exist in the literature{; most} are based on differences of the wall-normal velocity gradients, with others based on streamwise differences in flow variables~\cite{gul2022experimental}. Here, we follow the method of~\citeA{li2021experimental}, which takes streamwise differences in turbulence intensity, using the following equation:
\begin{equation}
    \Delta \Bigg[\frac{\langle u'u'\rangle(x,z)}{U^2_\infty}\Bigg] \Big{/} \Delta \Bigg[\log_{10}(\frac{\hat{x}}{\delta})\Bigg] \rightarrow 0.
    \label{eqn:delta_ibl}
\end{equation}

\noindent Equation \ref{eqn:delta_ibl} describes the difference {between two locations (given by the $\Delta$ operator)} of the normalized value of $\langle u'u' \rangle$, which is a function of streamwise ($x$) and wall-normal ($z$) coordinates, divided by the normalized distance between log-spaced streamwise stations. 
The value of $\delta_i$ at the upstream streamwise station is determined as the wall-normal height in which this difference approaches zero.  
Here, $U_\infty$ is the freestream velocity.
For our simulations we choose a threshold value of $10^{-4}$ to represent convergence toward zero in Equation~\ref{eqn:delta_ibl}.
{The} method {of}~\cite{li2021experimental} {was} compared against {other correlations for the same flow configuration} in~\citeA{gul2022experimental}, {and was found to perform favorably as it provided similar values to the other methods investigated.} 
        
\subsection{Quadrant Analysis}
\label{ssec:QA}
A good indicator of turbulence production is the Reynolds shear stress, $\langle u'w'\rangle$, and one method to understand its generation is quadrant analysis~\cite{wallace1972,willmarth1972structure}. 
{Here, we indicate a time-averaging with $\langle \cdot \rangle$.}
Quadrant analysis has been used to analyze smooth wall-bounded flows~\cite{wallace1972,willmarth1972structure}, rough wall-bounded flows~\cite{raupach1981conditional,choi1993direct,bristow2021unsteady}, {ABL flows}~\cite{lin1997effect,salesky2017nature,aksamit2018effect,li2019contrasts,rezaie2024characterizing}, and flows encountering a roughness transition~\cite{gul2022experimental}. 
A comprehensive review of quadrant analysis and its use is provided in~\citeA{wallace2016quadrant}.
The basic premise is to decompose the product of streamwise and wall-normal velocity fluctuations, $u'$ and $w'$, respectively, into four groups to understand the transfer of momentum. 
These groups are characterized as outward interactions ($Q_{1}$; $u' > 0$, $w' > 0$), ejections ($Q_{2}$; $u' < 0$, $w' > 0$), inward interactions ($Q_{3}$; $u' < 0$, $w' < 0$), and sweeps ($Q_{4}$; $u' > 0$, $w' < 0$)~\cite{wallace1972,wallace2016quadrant}. 
Under the background mean shear with $\partial U / \partial z > 0$ {$\langle u'w'\rangle$ is negative, and} $Q_{1}$ and $Q_{3}$ motions positively contribute, {decreasing the magnitude of} $\langle u'w'\rangle$; conversely, $Q_{2}$ and $Q_{4}$ motions augment {the magnitude of} $\langle u'w'\rangle$, acting as turbulence producing motions~\cite{wallace1972,willmarth1972structure}.
Contributions of each quadrant to the overall $\langle{u'w'}\rangle$ are considered to identify dominant motions in the flow. 
The contribution is calculated as
\begin{linenomath*}
\begin{equation}
    S_i = \frac{\sum^N_{k=1} u'w'_{i,k} I_i}{\sum^N_{k=1} u'w'_k},
    \label{eqn:quadContr}
\end{equation}
\end{linenomath*}

\noindent where $N$ is the total number of events observed for all quadrants, $i$ is the quadrant of interest ($i = 1,2,3,4$), $k$ is the event number, and $I_i = 1$ if {${u'w'}_{i,k}$} is located in the quadrant of interest and $I_i = 0$, otherwise. 

\subsubsection{Extension of Analysis through Time-Duration and Impulse}
\label{sssec:extQA}
A recent study conducted by~\citeA{bristow2021unsteady} extended the quadrant analysis over a barchan dune, by quantifying an average time-duration, $T_Q$, and average impulse strength, $J_Q$, for a quadrant event. 
The spatial evolutions of $J_Q$ and $T_Q$ provide insights into the impact of IBL development on turbulence producing motions and the sediment transport. 
{We first define a quadrant event using start and end times $A_{Q,k}$ and $B_{Q,k}$, respectively, where $Q$ is the quadrant number and $k$ is the event number.
$A_{Q,k}$ and $B_{Q,k}$ are found when the product of $u'$ and $w'$ cross zero, such that an event may start when $u'w' < 0 \rightarrow u'w' > 0$ and ends when $u'w' > 0 \rightarrow u'w' < 0$, or \textit{vice versa}.}
{Using this definition, we can determine the} average time-duration for quadrant events, as provided in~\citeA{bristow2021unsteady}, 
\begin{linenomath*}
\begin{equation}
    T_Q(x,z) = \frac{1}{N}\Bigg[\sum^N_{k=1} (B_{Q,k} - A_{Q,k}) \Bigg].
    \label{eqn:TQ}
\end{equation}
\end{linenomath*}

\noindent The average time-duration can be re-written in non-dimensional form as
\begin{linenomath*}
\begin{equation}
    T^*_Q = \frac{T_Q U_c}{H_c} \equiv \frac{T_Q U_\infty}{k_a},
    \label{eqn:TQstar}
\end{equation}
\end{linenomath*}

\noindent where $U_c$ is a characteristic velocity and $H_c$ is a characteristic length scale, that together form a characteristic duration $H_c$/$U_c$ ~\cite{bristow2021unsteady}.
We select the free stream velocity, $U_\infty$, as our characteristic velocity, and the average dune height, $k_a$, from~\citeA{gunn2021circadian} as our characteristic length scale.
The impulse of a quadrant event is taken as the integral of $\langle{u'w'}\rangle$ over the event duration, providing insight into the potential for sediment transport~\cite{bristow2021unsteady}.
As in~\citeA{bristow2021unsteady}, this integral is found with
\begin{linenomath*}
\begin{equation}
    J_Q(x,z) = \frac{1}{N}\sum^N_{k=1}\Bigg[\int^{B_{Q,k}}_{A_{Q,k}} \langle{u'(x,z,t)w'(x,z,t)}\rangle dt\Bigg].
    \label{eqn:JQ}
\end{equation}
\end{linenomath*}

\noindent $J_Q$ can be written in a non-dimensional form using a characteristic duration ($H_c/U_c$), and a velocity magnitude~(\citeA{bristow2021unsteady}). 
We use the same characteristic duration as before, and an average friction velocity over the dune field, $u_{\tau,02}$.

\begin{linenomath*}
\begin{equation}
    J^*_Q = \frac{J_Q U_c}{H_c u^2_\tau} \equiv \frac{J_Q U_\infty}{k_a u^2_{\tau,02}}.
    \label{eqn:JQStar}
\end{equation}
\end{linenomath*}

\section{Numerical Setup and Methods}
\label{sec:numMethods}

The {set-up of the} simulations employed in this study {is} the same as {that} in previous work by the authors~\cite{cooke2024mesoscale}. 
{We emphasize that these are new calculations which were conducted with the purpose of collecting new data within the same computational setup} allowing us to examine new quantities, and conduct a new analysis on the same flow.
This study builds on some of our previous findings; specifically, the relation between the development of the IBL and the observed thickening of the Reynolds shear stress profile after the roughness transition. In this study we examine the turbulent motions that underpin the observed self similarity.
We begin by describing in detail the solver used, the data we validate against, and the studies conducted to ensure the results are insensitive to domain dimensions and grid spacing.

\subsection{Numerical Solver Details}
\label{ssec:numSolver}

We deploy the LES flow solver \textit{CharLES}, from Cadence Design Systems (Cascade Technologies), to simulate a neutrally buoyant atmospheric boundary layer flow. 
\textit{CharLES} is an unstructured grid, body-fitted, finite-volume flow solver that solves the filtered variable-density Navier-Stokes equations, in a low-Mach isentropic formulation~\cite{ambo2020aerodynamic,bres2023aeroacoustic}.
{Coriolis forces are not included as the Rossby number is much greater than unity.}
The code uses a second-order central discretization in space, and a second-order implicit time-advancement scheme~\cite{bres2023aeroacoustic}.
{Although the validity of second-order finite-volume flow solvers has recently been challenged for its use in ABL flows}~\cite{giacomini2020suitability}, {recent work from} \citeA{hochschild2024comparison}, {supports the ability for this particular solver. Their work compared the LES-results of the Space Needle within an ABL flow to wind and pressure measurements from its roof in a prior observation campaign, and found good agreement between measured and simulated turbulence intensities and spectra.}
It is written in C++ and uses message-passing-interface to allow for parallelization. 

The streamwise, spanwise, and wall-normal directions are represented by $x, y,$ and $z$, respectively, with instantaneous velocity directions $U, V,$ and $W$ 
(or $u_1$, $u_2$,  and $u_3$). 
The filtered equations of mass and momentum are given by

\begin{linenomath*}
\begin{equation}
    \frac{\partial\tilde{\rho}}{\partial{t}} + \frac{\partial\tilde{\rho}\tilde{u_i}}{\partial{x_i}} = 0,
    \label{eqn:mass}
\end{equation}
\end{linenomath*}

\begin{linenomath*}
\begin{equation}
    \frac{\partial\tilde{\rho}\tilde{u}_i}{\partial{t}} + \frac{\partial\tilde{\rho}\tilde{u}_i\tilde{u}_j}{\partial{x_j}} = -\frac{\partial\tilde{p}}{\partial{x_i}} + \frac{\partial}{\partial{x_j}}\Bigg[(\mu + \mu_{SGS})\bigg(\frac{\partial\tilde{u}_j}{\partial{x_i}} + \frac{\partial\tilde{u}_i}{\partial{x_j}} - \frac{2}{3}\delta_{ij}\frac{\partial\tilde{u}_k}{\partial{x_k}}\bigg) \Bigg].
    \label{eqn:mmntm}
\end{equation}
\end{linenomath*}

\noindent Here, $\tilde{\rho}$ is the density of the fluid, $\tilde{p}$ is the pressure, $\mu$ is the dynamic viscosity of the fluid, and $\mu_{SGS}$ is the subgrid scale (SGS) viscosity.
Filtered quantities are denoted with a $\tilde{\cdot}$.
The filter width is set by the grid-spacing, $\Delta$, with turbulent motions larger than $\Delta$ resolved, and those smaller parameterized with an SGS model. 
For the remainder of the paper, all quantities without $\tilde{\cdot}$ are assumed to be filtered.
The static-coefficient Vreman SGS model is used to close the SGS viscosity: \cite{vreman2004sgs}, 
\begin{linenomath*}
    \begin{equation}
    \mu_{SGS} = \tilde{\rho}C_V\sqrt{\frac{B_\beta}{\alpha_{ij}\alpha_{ij}}},
    \end{equation}
\end{linenomath*}

\noindent where $C_V$ is the Vreman coefficient, $\alpha_{ij}$ is the filtered velocity gradient, and $B_\beta$ is the second invariant of $\beta_{ij} = \Delta^2\alpha_{ij}\alpha_{ij}$. {The static Vreman model has been successfully deployed in prior ABL studies}~\cite{hwang2022large,hwang2023large,gorle2023investigation,hochschild2024comparison,oh2024large}; {however, we include a sensitivity study to a dynamic local Smagorinsky SGS model~\cite{goc2024wind} in the Supplemental Material, where we find insignificant differences in the results.}
The low-Mach equation of state is given by 
\begin{linenomath*}
    \begin{equation}
        \tilde{\rho} = \frac{1}{c^2}(\tilde{p} - p_{ref}) + \rho_{ref},
    \end{equation}
\end{linenomath*}

\noindent with the reference pressure and density, $p_{ref}$ and $\rho_{ref}$, respectively, and speed of sound of the fluid, $c$.  
In assuming an isentropic approximation of the flow, the formulation provides the benefits of a variable density, compressible solver, while also removing the time-step restriction associated with low-Mach flows~\cite{bres2023aeroacoustic}, which are common to ABL flows~\cite{hwang2022large}.
For describing statistical quantities, the instantaneous velocity (e.g., $U$) is decomposed into its time-averaged ($\langle U\rangle$) and fluctuating ($u'$; about the mean) component: $U(x,y,z,t) = \langle U\rangle(x,y,z) + u'(x,y,z,t)$. 

CharLES uses an isotropic Voronoi meshing scheme, which provides means to generate a highly scalable, high-quality body-fitted mesh, suitable for complex, irregular geometries~\cite{hwang2022large,cooke2023numerical,bres2023aeroacoustic,hwang2023large}.
The mesh generation is fully parallelized and automated, allowing for production of a grid with O(10M) elements in O(1) minutes using tens of processors. 
For mesh generation, a far-field grid spacing is first specified, $\Delta_{FF}$, where this relative length-scale sets the refinement.
Subsequent mesh spacing is then determined by $\Delta_{FF}/2^n$, where $n$ is the desired number of refinement levels.
More details of the Voronoi meshing technique are provided in~\citeA{bres2023aeroacoustic}.

\subsection{White Sands Field Data}
\label{ssec:WhiteSands}

\begin{figure}[htb!]
    \centering
    \includegraphics[width=\textwidth]{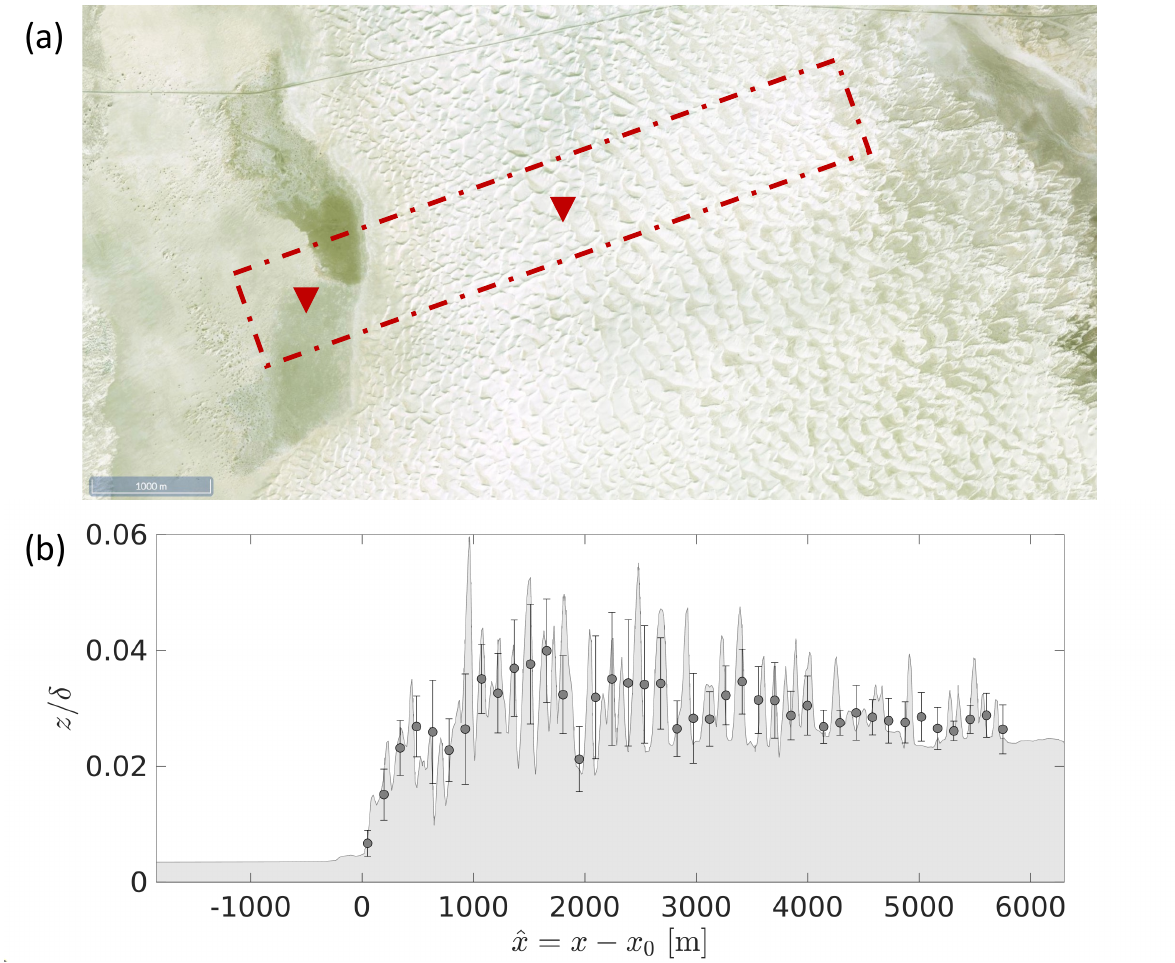}
    \caption{Field setting at White Sands dune field. (a) Satellite image of White Sands National Park, with the computational domain (red dot-dashed box) and two LiDAR observation stations (red triangles) marked. Flow is aligned with the dune formation direction.
    (b) A center line profile of the dune field in the numerical study (shaded area), featuring spanwise-averaged dune heights (gray circles), and their root-mean-square of the dunes along the span at each streamwise location (bars markers). All wall-normal elevations are normalized by $\delta = 300$ m.  }
    \label{fig:duneTopo}
\end{figure}

We investigate the smooth-to-rough surface transition found at White Sands National Park (Fig.~\ref{fig:duneTopo}a). 
The field data from White Sands included in this study have been extensively described in prior works~\cite{gunn2020macro, gunn2021circadian, cooke2024mesoscale}, and are briefly summarized here.
Open-source topographic data~\cite{usgs} for White Sands were gridded at 1 meter spatial ($x-y$) resolution, with vertical ($z$) resolution of $\sim 0.1$ m~\cite{gunn2020macro}.
The topography begins as a smooth playa surface (at $x = 0$ m), known as the Alkali Flat, and begins to rise  into a roughness transition region (at $x \approx 1.8$ km), where dunes form as low-amplitude ($\sim 10$ cm) sand waves with a fundamental wavelength $\sim 20$ m ~\cite{gadal2021spatial}.
Large transverse dunes abruptly emerge (at $x \approx 1.9$ km) and, around a kilometer after the transition, the transverse dunes break into isolated, heterogeneous barchan dunes whose migration rate and peak height decline gradually over several kilometers, after which the dunes are eventually immobilized by vegetation~\cite{jerolmack2012internal,reitz2010barchan,lee2019imprint}.

The average dune height across the dune field from prior studies was found to be $k_a = 3$ m~\cite{gunn2020macro,gunn2021circadian}. 
Because dune topography is spatially variable, however, we determine a spanwise-averaged dune elevation (in reference to the Alkali Flat $z =$ 0 m in the numerical domain) and the root-mean-square (rms) of this elevation, presented against a center line profile in Figure~\ref{fig:duneTopo}b.
Half kilometer bins of the dune field ($\hat{x} = 0-6$ km) are created to determine a localized $\hat{k}_a$ and root-mean-square, $\hat{k}_{rms}$.
Looking to Table~\ref{tab:duneHeight}, $\hat{k}_a$ increases over the first kilometer (Bins 1 and 2), reaching a maximum between $2-3.5$ km (Bins 3-5), then irregularly decreases over the remaining $\approx 3$ km of the dune field (Bins 6-12).
This behavior follows previously recorded and observed trends in dune height~\cite{jerolmack2012internal}, sediment flux~\cite{gunn2020macro}, and boundary stress~\cite{cooke2024mesoscale}.

\begin{table}[hbt!]
    \centering
    \caption{Half kilometer bins of the spanwise-averaged dune height and dune height root-mean-square at that streamwise location. Average heights are relative to the Alkali Flat ($z$ = 0 m). Corresponding bins for each streamwise probing stations are also indicated.}
    \begin{tabular}{l r r r}
    \hline 
       Bin & $\hat{x}$ Stations & $\hat{k}_a$ [m] & $\hat{k}_{rms}$ [m] \\
    \hline 
        1  & $\hat{x}_1$ and $\hat{x}_2$    & 5.3980     & 1.2508  \\
        2  & $\hat{x}_3$                    & 7.5215     & 2.3894  \\
        3  & $\hat{x}_4$                    & 10.4724    & 2.1155  \\
        4  & $\hat{x}_5$                    & 9.8482     & 2.3725  \\
        5  & $\hat{x}_6$                    & 10.1459    & 3.2984  \\
        6  & $\hat{x}_7$                    & 9.2489     & 2.2941  \\
        7  &                                & 9.5107     & 1.5359  \\
        8  & $\hat{x}_8$                    & 9.1641     & 1.6165  \\
        9  &                                & 8.3660     & 0.9759  \\
        10 & $\hat{x}_9$                    & 8.3996     & 1.0459  \\
        11 &                                & 8.2047     & 0.8953  \\
        12 &                                & 8.2867     & 1.2080  \\
    \hline 
    \end{tabular}
    \label{tab:duneHeight}
\end{table}

We use flow velocity data from the Field Aeolian Transport Events (FATE) campaign, which, using light detecting and ranging (LiDAR) equipment, collected horizontal and vertical velocity data from two fixed $x$ positions~\cite{gunn2021circadian}, shown in Figure~\ref{fig:duneTopo}a.
The LiDAR deployed in the study was a Campbell Scientific ZephIR 300 wind LiDAR velocimeter.
Initially, the equipment was placed on the Alkali Flat upwind of the dune field, where it collected vertical velocity profiles every 17 seconds over approximately 70 days during the spring windy season of 2017 at White Sands.
A year later, on the stoss side of a downstream dune, the LiDAR collected vertical velocity profiles every 17 seconds over approximately 25 days during the same windy season in 2018 at White Sands. 
Data were collected with a vertical resolution of 10 log-spaced bins, from $z$ = 10 m to $z$ = 300 m above the surface, with an additional point at $z$ = 36 m. 
In their study,~\citeA{gunn2021circadian} found that, {the effects of buoyancy at White Sands are highly transient and nonequilibrium, driven by an unsteady diurnal forcing, and} night-time winds produce a nocturnal jet that skims over a surface layer of cool stagnant air, which reduces boundary roughness effects and sediment transport. 
{A separate study from} \citeA{gunn2020macro} {used a longitudinal series of meteorological towers to examine downwind changes in flow velocity and sediment transport at White Sands and found that the effects of roughness were \textit{separable} from those of buoyancy.}
In order to {simulate a neutrally buoyant ABL and isolate the effects of roughness, we} remove the effects of buoyancy, by using only daytime measurements -- the twelve hour window from 06:00 to 18:00 local time.
The velocity within this window is then time-averaged to produce a daytime profile, and this process is completed for both the upstream and downstream LiDAR data.
The upstream profile is first used to derive the inflow conditions to the simulation, and later as validation for the inflow portion of the LES calculation.
Additionally,{in our previous work,} we{ validated the flow within the dune field (cf. their Fig. 1E)}~\cite{cooke2024mesoscale}.

\subsection{Computational Domain and Validation}
\label{ssec:CompDomainValid}

\begin{figure}[htb!]
    \centering
    \includegraphics[width=\linewidth]{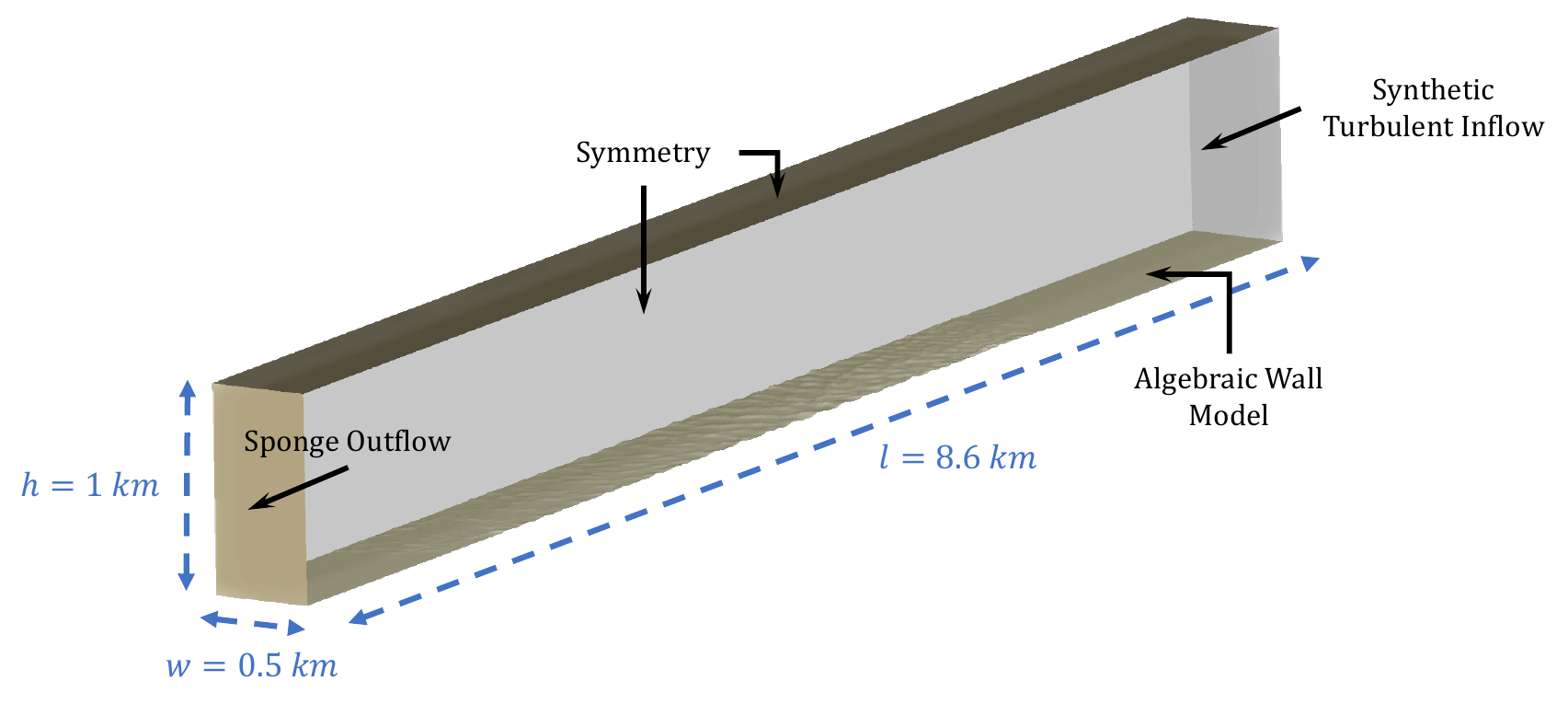}
    \caption{Simulation domain setup. A synthetic turbulent inflow boundary condition is imposed at the inlet, a numerical sponge is used at the outlet, the top and side walls deploy a slip boundary condition, and an algebraic wall-model is applied to the bottom surface, which is the scan of the dune field topography. The domain is $8.6$ km in streamwise length, $0.5$ km in spanwise width, and $1$ km in height.}
    \label{fig:fig2}
\end{figure}

We numerically analyze a neutrally-buoyant statistically steady ABL flow over an 8.6- by 0.5-km domain of the White Sands topographic data, depicted in Figure~\ref{fig:fig2}, oriented in the direction of dominant winds and dune migration ($\sim$ 15 degrees N of E).
The domain length is set to capture the mesoscopic scale of the IBL development, $\sim 30\delta$, where $\delta$ is the ABL height. 
Similarly, the width, determined in the sensitivity study {(See Supplemental for details)}, is much larger than an individual dune. 
At White Sands, prior observations of $\delta$ have estimated the thickness to fluctuate {in the daytime and nighttime} between {100-1000} m~\cite{norton1976diurnal}. 
We choose the height of the domain, $h$, to be {the upper limit of this range} $h = 1000$ m.
The height of the ABL is chosen to be $\delta = 300$ m, as this is the height of the highest data collection point in the FATE campaign~\cite{gunn2020macro}, preventing validation above this elevation, and it lies near the midpoint of the previously observed range at White Sands. 
For the top and sidewalls of the domain, we employ a symmetry boundary condition. 
{We were unable to apply periodicity to the side walls due to the uneven nature of the dune topography, and the topography could not be easily modified to allow for flat sides.}
At the inflow, a synthetic inflow generation based on digital filter techniques is implemented~\cite{klein2003digital}, which has been previously used in non-equilibrium WMLES studies~\cite{hu2023wm,hayat2023wm,cooke2024mesoscale}.
At the outflow a numerical sponge is deployed to minimize numerical effects~\cite{mani2012analysis,bodony2006analysis}, and the end of the domain is extended to establish a sponge zone that does not influence the study.
On the Alkali Flat and the dune field, we employ the algebraic (equilibrium) wall-model~\cite{bodart2011wall,kawai2012wall}, derived from the simplified boundary layer equations which assumes only the wall-normal diffusion. 

\begin{figure}[hbt!]
    \centering
    \includegraphics[width=\linewidth]{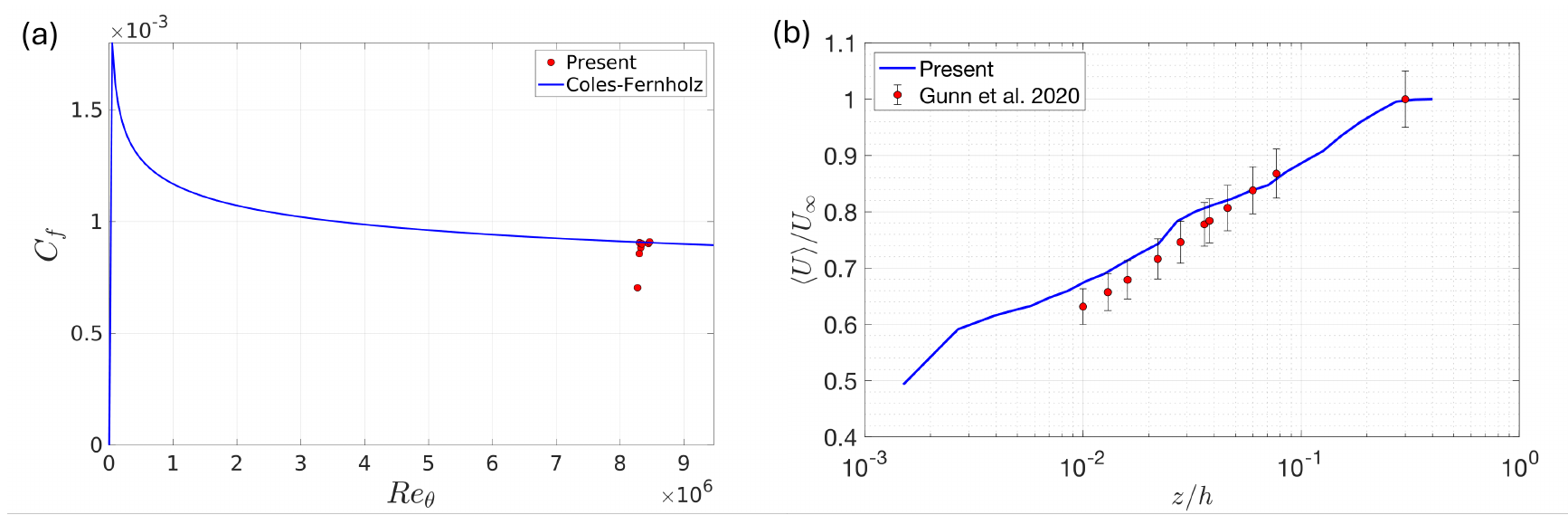}
    \caption{Inflow boundary condition characterization and validation.
    (a) Comparison of our simulated skin-friction coefficient as a function of $Re_\theta$ (black, red-filled circles) -- at multiple streamwise locations upstream of the roughness transition -- against the Coles-Fernholz correlation for $Re_\theta$~\cite{fernholz1996incomp} (blue line). (b) Validation of $\langle U\rangle$ from a coarse LES calculation (blue line) against the FATE campaign data~\cite{gunn2021circadian} (black, red-filled squares) with error bars showing $+/-$ 5\%. Values from the simulation are normalized by $U_\infty$ from the simulation, whereas values from the field campaign are normalized with an experimental $U_\infty$.}
    \label{fig:fig3}
\end{figure}

We extend the numerical domain of the Alkali Flat to first ensure the inflow profile becomes a fully-developed, zero-pressure-gradient turbulent boundary layer (ZPG TBL), before encountering the roughness transition.
As a check we examine the evolution of the skin-friction coefficient $C_f$ against  the momentum thickness Reynolds number $Re_\theta$.
As seen in Figure~\ref{fig:fig3}a, $C_f$ converges towards the empirical correlation for a ZPG TBL, well before the roughness transition, giving confidence to the inflow development.
Additionally, low-speed streaks closest to the wall are shown to have lengths of about  $2\delta_0-3\delta_0$, where $\delta_0 \approx 30$ m is the Atmospheric Surface Layer (ASL) height, further matching what would be expected of fully developed turbulent flow~\cite{hwang2016inner}.
As a test of validity for our model, we compare the time-averaged horizontal velocity profile of our simulation to the observations of FATE on the Alkali Flat, presented in Figure~\ref{fig:fig3}b.
The values from the simulation agree within 5\% error compared to the observed values. 
Considering that the field data are averaged over a non-stationary forcing, and that buoyancy effects are not included in the simulation, this agreement is {encouraging}, indicating that treatment of the ABL flow at White Stands as steady and neutrally buoyant is an appropriate model.

The mesh deployed in the study contained approximately $85\times10^6$ control volumes, and used $\Delta_{FF} = 28$ m with five levels of refinement, yielding a minimum grid-spacing $\Delta_{min} = 0.75$ m closest to the surface. 
We note that despite the refinement leading to the average dune height being resolved by four control volumes, in viscous units our near-wall spacing is $\Delta^+_{min} \approx 3200$, due to the high $Re_\tau$.
Here, $\cdot^+$ denotes scaling with viscous units, $\delta_\nu \equiv \nu/u_\tau$ and $u_\tau$.
However, we adequately resolve the ABL height with $\approx$ 67 control volumes, as well as the IBL with a minimum of $\approx$ 20 and a maximum of $\approx$ 46, over the course of its development. 
We next justify our mesh choice with a grid sensitivity study.

\section{Results}
\label{sec:results}

\subsection{IBL Development and Turbulence Within}
\label{ssec:IBLResult}

\begin{figure}[hbt!]
    \centering
    \includegraphics[width=\linewidth]{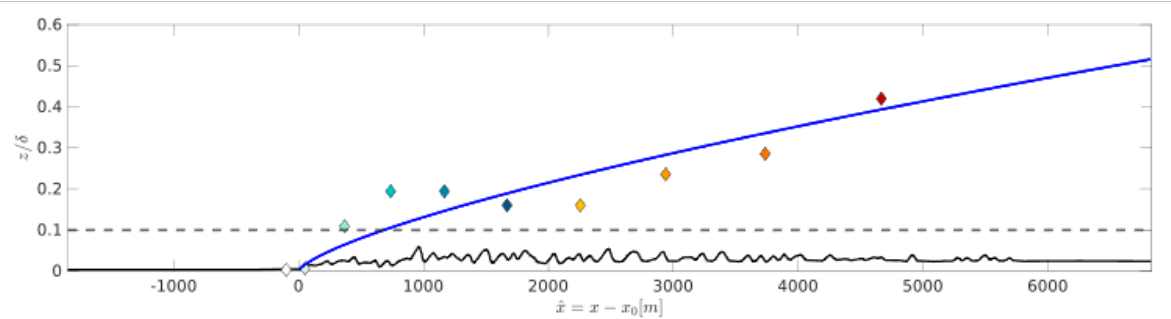}
    \caption {Calculated IBL height (multi-colored diamonds) along the center line of the dune field (black line). A correlation (blue line) for the data is found to be $\delta_i/z_{02}$ = $0.29(\hat{x}/z_{02})^{0.71}$, where $z_{02} = 10^{-1}$ m~\cite{gunn2020macro,gunn2021circadian}.}
    \label{fig:iblRSS}
\end{figure}

We first calculate the {IBL height} downstream of the roughness transition using Equation~\ref{eqn:delta_ibl}, at ten streamwise log-spaced stations from $\hat{x}_1 = 50$ m to $\hat{x}_{10} = 5751$ m (Fig.~\ref{fig:iblRSS}).
Wall-normal elevations are probed at equal intervals, between $0-400$ m, for $\langle u'u'\rangle$. 
We use the mean velocity at $z = 400$ m -- which is well outside the ABL -- for $U_\infty$ and $\delta$ to normalize $\hat{x}$.

\noindent A correlation of the form given in Equation~\ref{eqn:ibl_corr} is fit to the data, and we use $z_{02} = 10^{-1}$ m, found by~\citeA{gunn2020macro,gunn2021circadian} in the FATE campaign. 
We find the coefficient $a = 0.29$ and the exponent $b = 0.71$. 
Our correlation from the determined values of $\delta_i$ has good agreement with classic scaling models~\cite{elliott1958growth,townsend1965response,antonia1971response,wood1982internal,pendergrass1984dispersion,savelyev2001notes}.
Additionally, our power-law value, $b$, agrees well with~\citeA{li2021experimental}; using this method, they found values of $a = 0.75$ and $b = 0.77$ for their two datasets in their rough-to-smooth transition study.

We note that the correlation is constructed with an assumed homogeneous roughness parameter governing the whole dune field. 
Looking at Table~\ref{tab:duneHeight}, we see there is significant variability in $\hat{k}_a$ over the first few kilometers. 
Over the first kilometer of the dune field, relatively low amplitude dunes are superimposed on a topographic ramp,
Beyond this ramp, dunes grow in size while the underlying topography levels out (Fig. \ref{fig:duneTopo}b).
Thus, the flow may actually experience two roughness transitions; first, from the smooth Alkali Flat to the low-lying transverse dunes, and second from the low-lying dunes to the larger isolated barchan dunes.
Looking to Figure~\ref{fig:iblRSS}a, the simulated IBL heights are suggestive of two transitions -- or, at least, the IBL height exhibits long-wavelength fluctuations. Nevertheless, assuming homogeneity in roughness captures the first-order growth of the IBL. 

\subsection{The Role of $\delta_i$ in Turbulence within the IBL}
\label{ssec:IBLLengthScale}

\begin{figure}[hbt!]
    \centering
    \includegraphics[width=\linewidth]{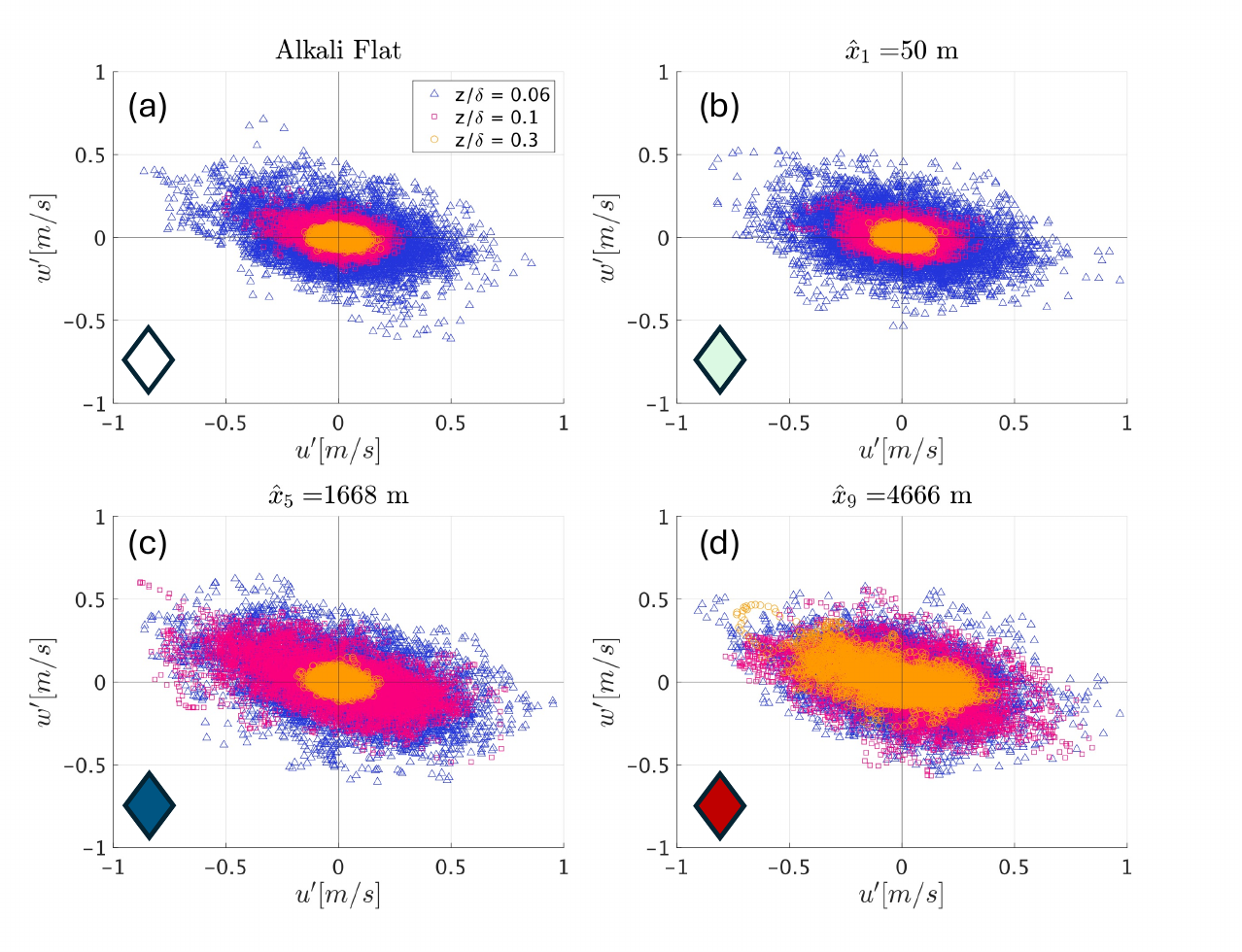}
    \caption{Quadrant analysis of turbulence at three relative bed heights, for four downstream probing stations. Diamond colors correspond to the location within the dune field and IBL height from Figure~\ref{fig:iblRSS}. (a) Scatter plot of the product of the streamwise and wall-normal velocity fluctuations, $u'$ and $w'$, respectively, in the Alkali Flat. Here, we see healthy turbulence at the lowest wall-normal elevation $z/\delta$ = 0.06, where a majority of the motions reside in the second (upper left) and fourth (lower right) quadrants. At the second highest elevation at $z/\delta$ = 0.1, we expect to see lower magnitudes of the fluctuations, and a decrease in frequency of motions in the $Q_2$ and $Q_4$ as turbulence decays further from the wall. At the highest elevation, $z/\delta$ = 0.3, we expect to see lower magnitudes of fluctuations, and no preference for $Q_2$ and $Q_4$. (b-d) Same plots as (a), at stations downstream of the roughness transition. Magnitude of fluctuations, and the frequency pf $Q_2$ and $Q_4$ motions, at $z/\delta$ = 0.1 increases with streamwise distance (c). For $z/\delta$ = 0.3, these quantities begin to increase further downstream (d). Overall, we see an increase in the magnitude and frequency of turbulence producing motions away from the wall, as the IBL grows to subsume these higher elevations.}
    \label{fig:scatter}
\end{figure}

As outlined in Section~\ref{ssec:QA}, turbulent momentum transport {is} associated with $\langle u'w'\rangle$, and may be quantified through quadrant analysis.
We plot $u'$ and $w'$ in the Alkali Flat and downstream in the dune field, at three wall-normal elevations: $z/\delta = 0.06, 0.1$ and $0.3$ (Fig.~\ref{fig:scatter}).
In the smooth Alkali Flat, we observe that the closest wall-normal location behaves as expected in typical ZPG TBLs (Fig.~\ref{fig:scatter}a). 
A majority of the points are located within the second and fourth quadrants, which represent the turbulence-producing ejection and sweep events, respectively. 
Moving away from the wall, the turbulence within the atmospheric surface layer is expected to gradually decrease. Indeed, at $z/\delta = 0.1$ the magnitude of the fluctuations is seen to decrease, with fewer points residing in $Q_2$ and $Q_4$.
Furthest from the wall, the fluctuating velocity magnitude is diminished, and the frequency of each motion is nearly indistinguishable; i.e., there is no observable preference for any quadrant.

Downstream of the roughness transition, the results from the lowest elevation remain largely unchanged (Fig.~\ref{fig:scatter}b-d).
Similarly, the results at $z/\delta = 0.1$ and $0.3$ at the first station (Fig.~\ref{fig:scatter}b) closely resemble those of the Alkali Flat, as the relevant length-scale is still $\delta_0$ here.
However, we begin to see changes to these wall-normal elevations as we move further downstream, and $\delta_i$ grows larger than $\delta_0$.
For $z/\delta = 0.1$, the change is {noticed once} $\delta_i > \delta_0$.
This is reflected by both an increase in the magnitude of the velocity fluctuations, and in the number of events in $Q_2$ and $Q_4$.
By $\hat{x}_5$ (Fig.~\ref{fig:scatter}{c}), the magnitude of the fluctuations at $z/\delta = 0.1$ matches those found at $z/\delta = 0.06$, and the frequency of sweep and ejection events is nearly equivalent; this behavior is maintained downstream.
Furthest from the wall, we observe almost no change in magnitude or frequency from the Alkali Flat, reflecting the dissipation of the Reynolds shear stress, and subsequently, the turbulent motions, further from the wall beyond $\delta_0$. 
However, {by the final station} (Fig.~\ref{fig:scatter}{d}), there is a clear change to the flow, as the IBL height {has surpassed} $z/\delta = 0.3$; this is evident from the increase in both magnitude and frequency of ejection events.
The enhancement of ejection events is associated with $\delta_i$ growing beyond $z/\delta = 0.3$. The upshot of these observations is this: as the IBL thickens downwind, regions of the flow that are subsumed by it exhibit increased turbulence producing motions.

\begin{figure}[hbt!]
    \centering
    \includegraphics[width=\linewidth]{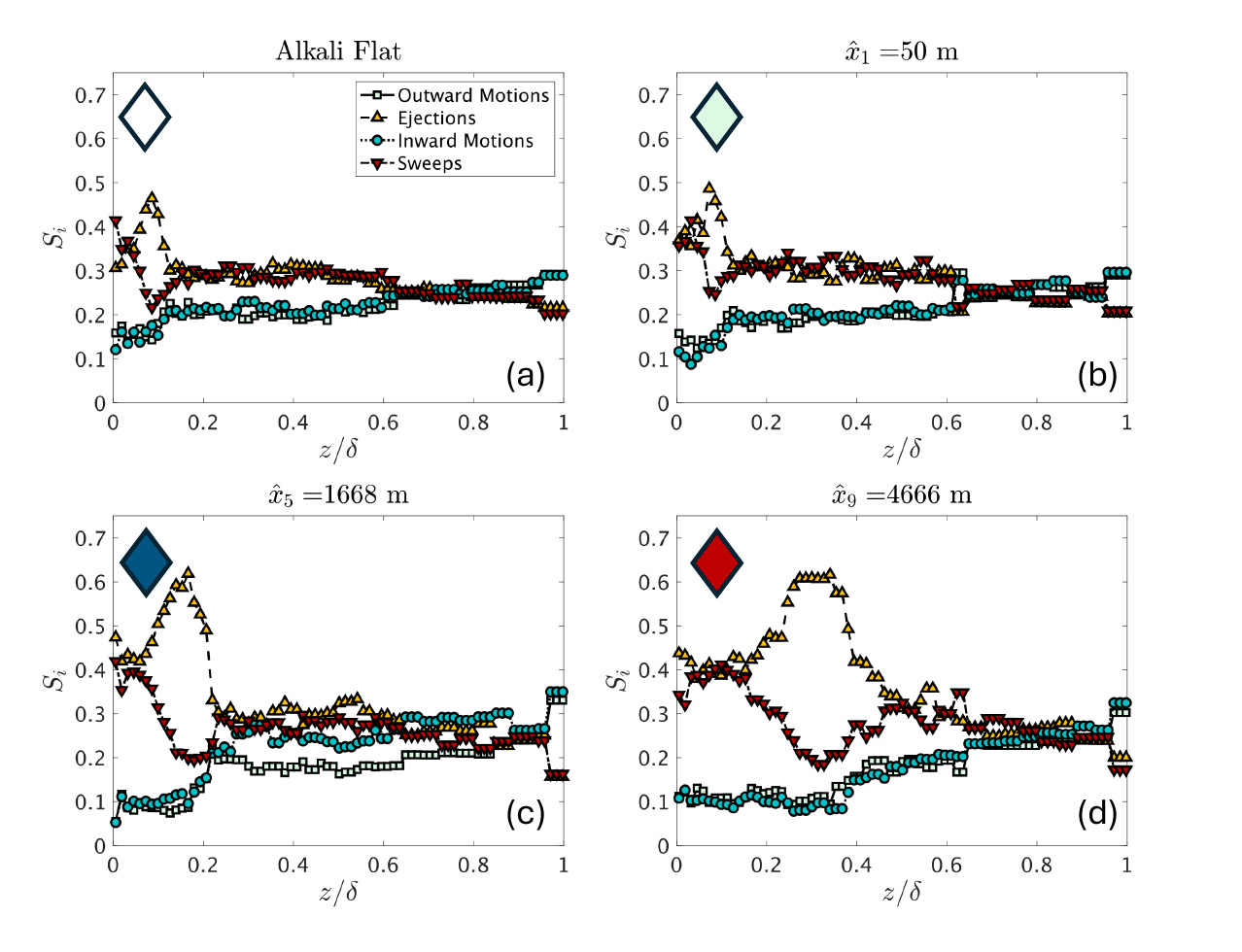}
    \caption{Contributions to the Reynolds shear stress, $S_i$, where $i = 1, 2, 3,$ and $4$, corresponding to each quadrant. Diamond colors correspond to the location within the dune field and IBL height from Figure~\ref{fig:iblRSS}.  
    (a) The Alkali Flat station. Near the bed, $Q_2$ and $Q_4$ contributions are roughly equal. Farther from the wall the contribution from ejections increases, with a corresponding decrease in contribution from sweeps. Farther still the contributions become roughly equal again. (b-d) Same plots as (a), at stations downstream of the roughness transition. We observe a systematic downstream increase in the distance from the wall where $Q_2$ and $Q_4$ motions contribute to $\langle u'w'\rangle$. 
    }
    \label{fig:quadContr}
\end{figure}

With Equation~\ref{eqn:quadContr}, we quantify the contributions to the Reynolds shear stress by each quadrant, $S_i$, and observe the changes occurring downstream of roughness transition.
In the Alkali Flat (Fig.~\ref{fig:quadContr}a) closest to the surface, the majority of the contributions to $\langle u'w'\rangle$ come from $Q_4$ events. Farther from the wall, however, $Q_2$ motions become the highest contributor and $Q_4$ contributions correspondingly decrease. 
At $\approx z/\delta = 0.1 \equiv \delta_0$ the contributions from $Q_2$ and $Q_4$ events become nearly equivalent. By $\approx z/\delta = 0.6$, all quadrant motions have approximately equal contribution.
The trends are similar at $\hat{x}_1$ (Fig.~\ref{fig:quadContr}b). We see {changes downstream once the IBL has grown} (Fig.~\ref{fig:quadContr}c,d), where the local spike in $Q_2$ contributions (and corresponding dip in $Q_4$ contributions) moves farther from the wall. As we continue to move downstream, the associated spike (and dip) systematically moves to higher relative heights above the wall.
there is an increase in distance from the surface where the $Q_2$ contribution peaks.
This corresponds as well to a distance further from the wall where contributions from $Q_2$ and $Q_4$ events becomes nearly equal again. 
This perceived trend continues downstream, where the $Q_2$ and $Q_4$ contributions are initially similar, then diverge with peak contributions from $Q_2$ motions occurring further from the wall with increasing $\hat{x}$. 

\begin{figure}[hbt!]
    \centering
    \includegraphics[width=\linewidth]{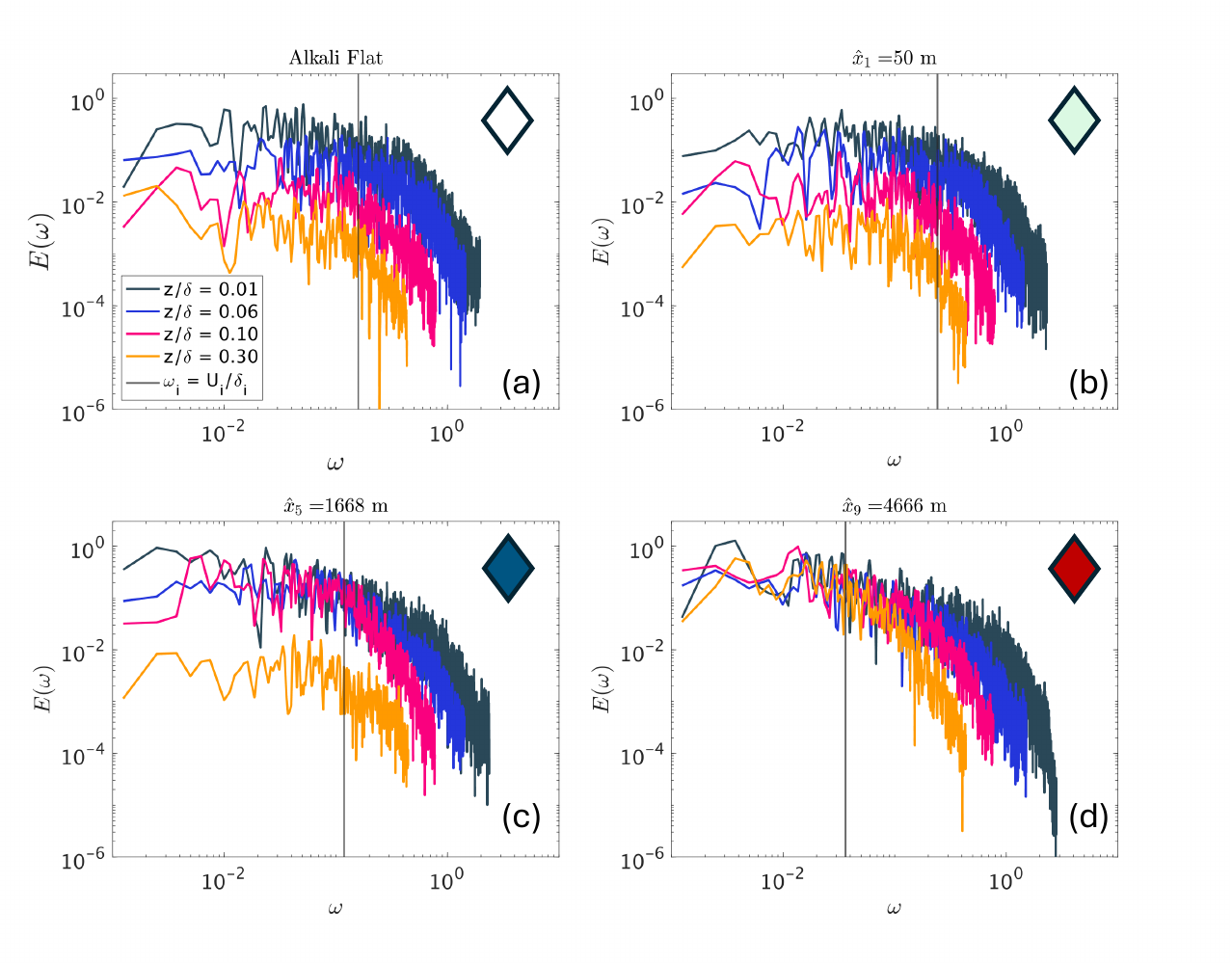}
    \caption{Energy frequency spectrum of the streamwise velocity fluctuation, $E(\omega)$, at four heights for the ten probing stations. 
    Diamond colors correspond to the location within the dune field and IBL height from Figure~\ref{fig:iblRSS}. Vertical black line represents the frequency associated with the eddy turnover time at the scale of the IBL, $\omega_i \equiv U_\infty(z$=$\hat{\delta})/\hat{\delta}$. 
    This frequency decreases as the IBL thickness grows, and corresponds roughly to the scale break in the energy spectra. 
    (a) Alkali Flat. $E(\omega)$ decreases with increasing distance from the wall, at all frequencies.  (b-d) Same plots as (a), at stations downstream of the roughness transition. We observe that as the IBL grows to subsume higher elevations, the magnitude of $E(\omega)$ for low-frequency turbulence grows. Note that once all elevations are within the IBL (j), the turbulent energy at low frequencies is the same for all elevations. 
    }
    \label{fig:PSD}
\end{figure}

To uncover the downstream evolution of the length- and time-scales in the flow, we examine changes to the energy frequency spectrum of the the streamwise velocity fluctuations, $E(\omega)$ (Fig.~\ref{fig:PSD}).
{Note, we sample the data with the same frequency, and thus the output frequency range is the same for the raw data; however, the maximum value of the frequency we include in the analysis decreases due to the grid resolution decreasing away from the wall.}
We take the long time-series data at four wall-normal elevations: $z/\delta = 0.01, 0.06, 0.1,$ and $0.3$ and transform the data into the frequency domain using the Fourier transform, in order to compute $E(\omega)$.
Further details of this process are given in~\cite{park2016space}.
Upwind of the dune field, $E(\omega)$ in the Alkali Flat (Fig.~\ref{fig:PSD}a) demonstrates the reduction in turbulence further from the wall, as the energy contained in the flow at all frequencies decreases. 
This trend is seen at $\hat{x}_1$ as well (Fig.~\ref{fig:PSD}b).
By {$\hat{x}_5$} (Fig.~\ref{fig:PSD}c), there is an observed increase in the energy contained at lower frequencies (larger scales) for $z/\delta = 0.06$ and $0.1$, such that their magnitudes are nearly equal. 
This observation is another sign of the influence of IBL growth; as $\delta_i$ grows to subsume these elevations, we see an increase in large-scale turbulence.
Further from the wall, at $z/\delta = 0.3$, values of $E(\omega)$ remain lower and unchanged as the IBL height has yet to reach this height. 
By {the final station, $\hat{x}_9$,} (Fig.~\ref{fig:PSD}{d),} the energy at all elevations for lower frequency turbulence is roughly equal. 
Within the dune field, we can compute the expected frequency of turbulent motions (using the frozen turbulence hypothesis) associated with the the scale of the IBL using $\omega = U_c/\hat{\delta}$, where $U_c \equiv \langle U\rangle(z = \hat{\delta})$.
With increasing downstream distance, the computed frequency associated with IBL-scale turbulence decreases. 
This computed frequency also corresponds roughly with the scaling break in the energy frequency spectrum, which is typically associated with the turnover time of the largest-scale eddies in the system. Taken together, results suggest that the developing IBL sets the scales of large-scale turbulence within it. 

\subsection{Evolution of $T_Q$ and $J_Q$}
\label{ssec:EvoTurbMotions}

We now focus on changes to the strength and time-duration for all quadrant events, with an emphasis on ejections and sweeps, after the roughness transition.
We use Equations~\ref{eqn:TQ} and \ref{eqn:JQ} to examine how turbulence producing motions evolve after the roughness transition.
We collect time-series data for over $50$ large-eddy turnover times, $T \equiv \delta/U_\infty$, resulting in $\sim O(10^3)$ separate quadrant events detected at each elevation and streamwise location.
We record events at all streamwise stations in the dune field up to $z/\delta = 1.33$, and will focus on four elevations: $z/\delta = 0.01, 0.06, 0.1,$ and $0.3$.
We include data at the Alkali Flat to provide a baseline to compare against.
As was done by~\citeA{bristow2021unsteady}, we do not filter events to prevent loss of long-time quadrant events that have smaller magnitudes.

\begin{figure}[hbt!]
    \centering
    \includegraphics[width=\linewidth]{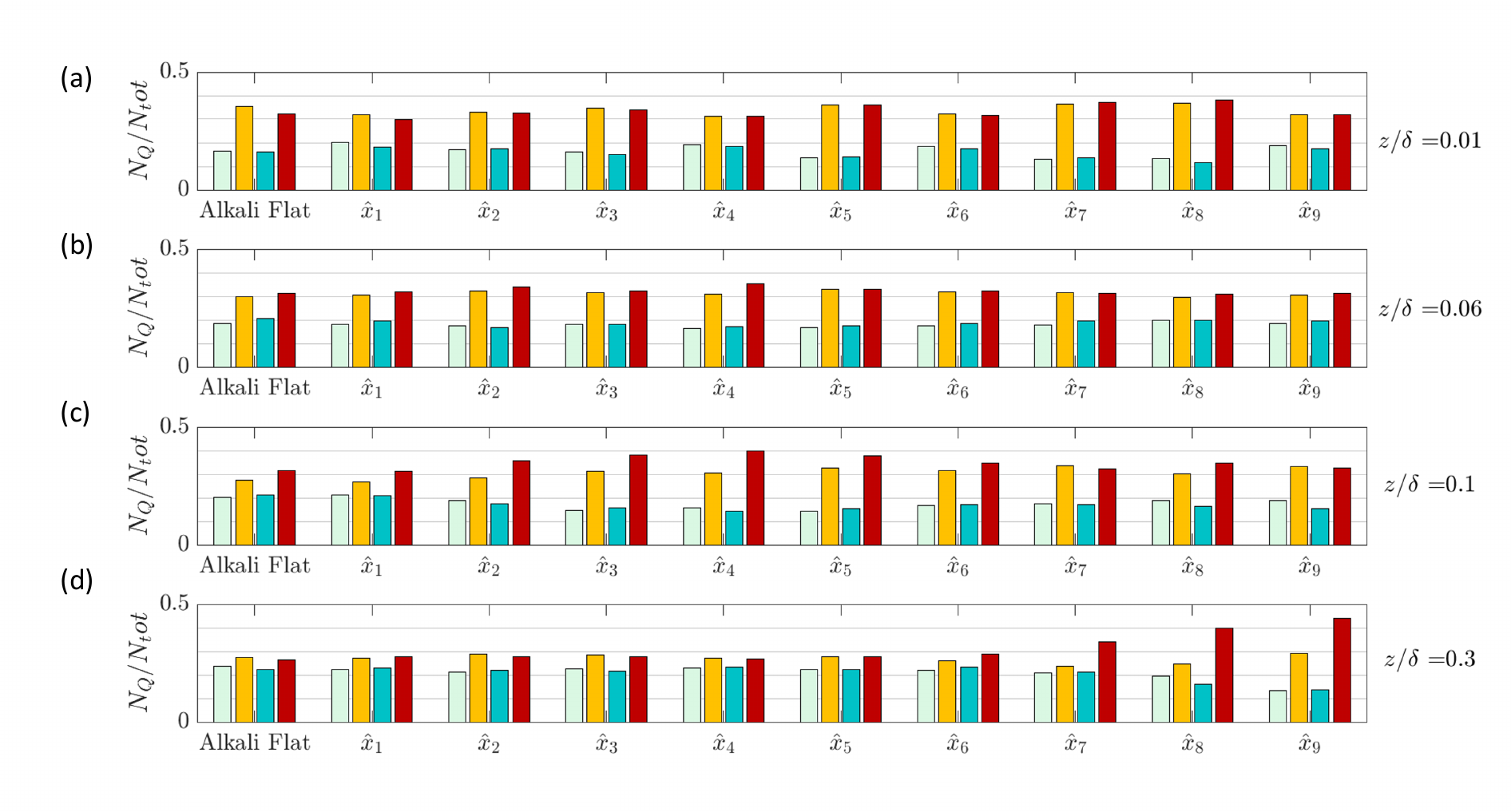}
    \caption{Relative frequency of quadrant events for each probing station, at distinct relative heights  $z/\delta$. Bars are in order of quadrant (left to right), with mint-green for $Q_{1}$, yellow for $Q_{2}$, blue for $Q_{3}$, and red for $Q_{4}$.  (a) The lowest elevation $z/\delta$. Behavior is as expected, with $Q_2$ and $Q_4$ events occurring most frequently. (b) $z/\delta = 0.06$, (c) $z/\delta = 0.1$, and (d) $z/\delta = 0.3$. Note the relative increase in $Q_2$ and $Q_4$ events at higher elevations moving downwind; as the IBL grows to subsume higher elevations, these turbulence producing motions increase.}
    \label{fig:freqOfEvents}
\end{figure}

To understand changes to the frequency of events in the streamwise direction at each elevation, we take the ratio of the number of events in one quadrant to the total number of events in all quadrants (Fig.~\ref{fig:freqOfEvents}).
Nearest the surface  at $z/\delta = 0.01$ (Fig.~\ref{fig:freqOfEvents}a), the frequency of events looks as expected in both the Alkali Flat and the dune field, with a majority of the events being ejections and sweeps.
Looking first at results for the Alkali Flat further from the wall, at $z/\delta = 0.06$ (Fig.~\ref{fig:freqOfEvents}b), there is a slight reduction in $Q_2$ and $Q_4$ events, and a concomitant increase in $Q_1$ and $Q_3$ events.
At $z/\delta = 0.1$ (Fig.~\ref{fig:freqOfEvents}c) the frequencies of each event type change little from $z/\delta = 0.06$. 
However, farthest from the wall at $z/\delta = 0.3$ (Fig.~\ref{fig:freqOfEvents}d), we see the frequency of $Q_1$ and $Q_3$ events has increased, causing an additional decrease in $Q_2$ and $Q_4$ events. 
This is expected as the turbulence producing motions were seen to decrease away from the wall in the Alkali Flat (Fig. ~\ref{fig:scatter}a).
Focusing now on changes within the dune field, at the lowest elevation (Fig.~\ref{fig:freqOfEvents}a), there are subtle differences between $\hat{x}_5$ and $\hat{x}_8$, where there is a definite reduction in the $Q_{1}$ and $Q_{3}$ events and concomitant increase in the frequency of $Q_{2}$ and $Q_{4}$ events. 
At $z/\delta =$ 0.06 (Fig.~\ref{fig:freqOfEvents}b), we find at all $\hat{x}$ that $Q_{2}$ and $Q_{4}$ events occur more frequently, although the highest (lowest) frequency of $Q_{2}$ and $Q_{4}$ ($Q_{1}$ and $Q_{3}$) events occurs between $\hat{x}_4$ to $\hat{x}_6$. 
At $z/\delta = 0.1$ (Fig.~\ref{fig:freqOfEvents}c), event frequency at $\hat{x}_1$ is similar to the Alkali Flat at the same elevation. There is a significant change, however, beginning at $\hat{x}_2$.
Since the IBL height has surpassed the ASL height, there is an observed increase in $Q_2$ and $Q_4$ event frequency, continuing for all other $\hat{x}$. 
Farthest from the wall (Fig.~\ref{fig:freqOfEvents}d), event frequency for all quadrants from $\hat{x}_1$ to $\hat{x}_6$ is similar to the Alkali Flat.
Beginning at $\hat{x}_7$, $Q_4$ event frequency increases, and continues to do so further downstream.
We observe a concomitant decrease in $Q_{1}$ and $Q_{3}$ events at these stations. These observed changes in event frequency are due to the IBL height surpassing $z/\delta = 0.3$.

\begin{figure}[hbt!]
    \centering
    \includegraphics[width=\linewidth]{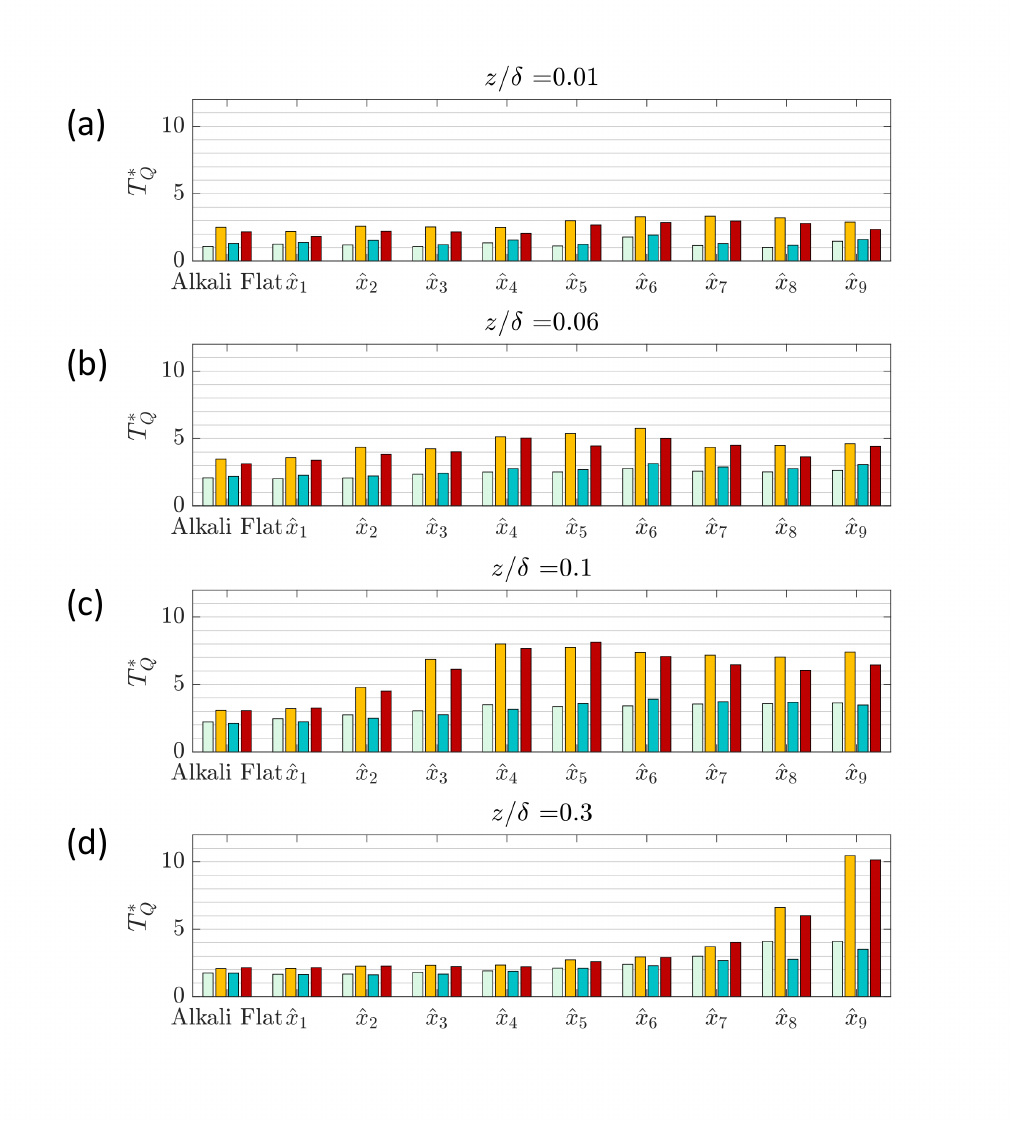}
    \caption{Average dimensionless time duration of quadrant events, $T^*_Q$, at each probing station, at distinct relative heights $z/\delta$. Bars and colors are the same as in Figure~\ref{fig:freqOfEvents} (a) $z/\delta$ = 0.01, (b) $z/\delta$ = 0.06, (c) $z/\delta$ = 0.1, and (d) $z/\delta$ = 0.3.}
    \label{fig:TQ}
\end{figure}

We present the results of the average time-duration, $T_Q$, for each quadrant event, given in its non-dimensionalized form, $T^*_Q$ (Equation~\ref{eqn:TQstar}).
Results for the Alkali Flat and all $\hat{x}$ in the dune field are presented in Figure~\ref{fig:TQ}.
For calculating $T^*_Q$, we choose $U_c$ to be $U_\infty$ (as before $\langle U\rangle(z = 400$ m)), and $k_a$ from the FATE campaign.
We first observe changes in elevation at the Alkali Flat.
Starting at $z/\delta = 0.01$ (Fig.~\ref{fig:TQ}a), we observe that $Q_2$ and $Q_4$ events are on average longer than $Q_1$ and $Q_3$ motions. 
For $z/\delta = 0.06$ (Fig.~\ref{fig:TQ}b), this pattern still holds; however, there is an increase in $T^*_Q$ for all events, including $Q_1$ and $Q_3$ motions.
This increase does not continue at higher elevations $z/\delta = 0.1$ and $0.3$ (Figs.~\ref{fig:TQ}c and~\ref{fig:TQ}d); the opposite trend is observed, where $T^*_Q$ is shown to decrease at higher elevations in the Alkali Flat. 
Within the dune field, the most distinctive pattern is that, for the two elevations closest to the wall (Figs.~\ref{fig:TQ}a and~\ref{fig:TQ}b), events corresponding to $Q_{2}$ and $Q_{4}$ are longer than those corresponding to $Q_{1}$ and $Q_{3}$ motions. 
Additionally, as was seen with the Alkali Flat, there is an increase in $T^*_Q$ magnitude when moving away from the wall. 
We observe a similar pattern as was seen with the frequency in Figure~\ref{fig:freqOfEvents}b, where the longest average events occur at the point where they have the highest frequency at $z/\delta = 0.06$ ($\hat{x}_4$ to $\hat{x}_6$).
For $z/\delta = 0.1$ (Fig.~\ref{fig:TQ}c), $T^*_Q$ at $\hat{x}_1$ has nearly identical values to those in the Alkali Flat, but by $\hat{x}_2$, there is a significant increase in $T^*_Q$, as the IBL height has surpassed the ASL height. 
This also holds for $z/\delta = 0.3$; $T^*_Q$ is similar at all $\hat{x}$ locations to that in the Alkali Flat, until $\hat{x}_7$, where the IBL height begins to surpass $z/\delta = 0.3$.
Outside of the IBL, $T^*_Q$ is mostly similar amongst all quadrants, with $Q_{2}$ and $Q_{4}$ retaining marginally larger values.
{To test the statistical significance of the increase in $T^*_Q$ -- for $Q_{2}$ and $Q_{4}$ -- within the IBL, we conduct a right-tail t-test ($\alpha = 0.05$). 
For $z/\delta = 0.01$, there is no significant increase in average time-duration for any quadrant at any point downstream of the transition. 
However, for $z/\delta = 0.06$, there is a significant increase in $T^*_Q$ for $Q_2$ and $Q_4$ at $\hat{x}_2$, and then for all quadrants downstream of $\hat{x}_3$.
The observations of $z/\delta = 0.06$ are the same at $z/\delta = 0.1$, but now all quadrants have a significant increase by $\hat{x}_2$.
For the highest elevation, $z/\delta = 0.3$, there is no significant increase in average time-duration until $\hat{x}_5$, where all quadrants see an increase, and continues to be the case downstream.
}





\begin{figure}[hbt!]
    \centering
    \includegraphics[width=\linewidth]{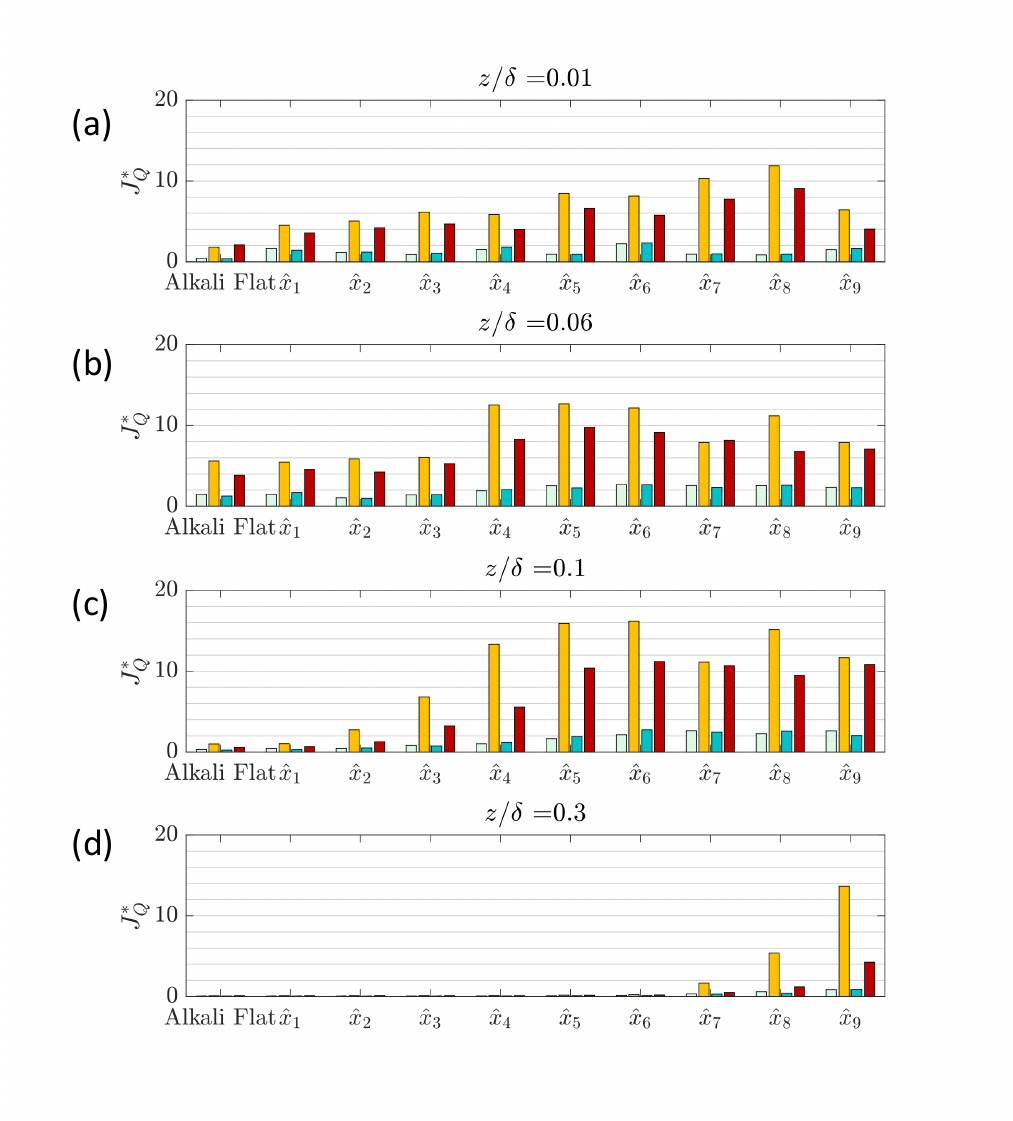}
    \caption{Average impulse strength of quadrant events, $J^*_Q$, for each probing station, at distinct relative heights $z/\delta$. Bars and colors are the same as in Figure~\ref{fig:freqOfEvents}. (a) $z/\delta$ = 0.01, (b) $z/\delta$ = 0.06, (c) $z/\delta$ = 0.1, and (d) $z/\delta$ = 0.3.
    }
    \label{fig:JQ}
\end{figure}

Lastly, we analyze the average event impulse strength, $J_Q$, in non-dimensional form, $J^*_Q$ (Equation~\ref{eqn:JQStar}). 
We use the same values for $U_\infty$ and $k_a$ as was used for $T^*_Q$, and additionally the average friction velocity over the whole dune field, $u_{\tau,02}$.
We begin by first looking at the evolution in the Alkali Flat.
Beginning with $z/\delta = 0.01$ (Fig.~\ref{fig:JQ}a), where, despite the low magnitude, it is clear that $Q_2$ and $Q_4$ events have higher average impulse values compared to $Q_1$ and $Q_3$ events.
Further, at $z/\delta = 0.06$ (Fig.~\ref{fig:JQ}b), the magnitude of $J^*_Q$ has significantly increased, especially for $Q_2$ and $Q_4$ events, but as was seen with $T^*_Q$, at the ASL height (Fig.~\ref{fig:JQ}c) and above (Fig.~\ref{fig:JQ}d), the average impulse decreases for all events. 
Much like $T^*_Q$, $J^*_Q$ follows similar trends observed within the dune field too.
For $z/\delta = 0.01$ (Fig.~\ref{fig:JQ}a), $J^*_Q$ maintains larger values for $Q_{2}$ and $Q_{4}$ events, compared to $Q_{1}$ and $Q_{3}$ events, at all $\hat{x}$.
Moreover, just as was observed with frequency of events and $T^*_Q$, the largest $J^*_Q$ values for $Q_2$ and $Q_4$ events at this elevation occurs between $\hat{x}_5$ to $\hat{x}_8$.
Moving to $z/\delta = 0.06$ (Fig.~\ref{fig:JQ}b), $Q_2$ and $Q_4$ events continue to display larger values of $J^*_Q$, with $Q_{2}$ events consistently having the highest magnitudes. 
Similar to $T^*_Q$, the largest values for $J^*_Q$ are found between $\hat{x}_4$ and $\hat{x}_6$.
Next, at $z/\delta = 0.1$ (Fig.~\ref{fig:JQ}c), values of $J^*_Q$ at $\hat{x}_1$ mimic those found in the Alkali Flat, and then begin to increase at $\hat{x}_2$ and $\hat{x}_3$. 
Again, the largest values for $J^*_Q$ at this elevation are also contained within $\hat{x}_4$ and $\hat{x}_6$.
Finally, at $z/\delta = 0.3$ (Fig.~\ref{fig:JQ}d), the magnitude of $J^*_Q$ is very low, until $\hat{x}_7$ where the IBL height begins to surpass $z/\delta = 0.3$.
Additionally, we observe a large disparity between the magnitudes of average impulse for $Q_2$ and $Q_4$ events.
{As was done for $T^*_Q$, we use a right-tail t-test ($\alpha = 0.05$) to test the statistical significance of the increase in $J^*Q$, downstream of the roughness within the IBL. 
At $z/\delta = 0.01$, there is a significant increase at all stations downstream of the transition.
For $z/\delta = 0.06$, interestingly, the significance is not seen until $\hat{x}_4$, whereas for $T^*_Q$, $Q_2$ and $Q_4$ saw a significant increase at $\hat{x}_2$. 
We observe from Table}~\ref{tab:duneHeight} {that the largest dunes are found at this station, indicating that perhaps closer to the wall the dunes hold a larger influence over $J^*Q$.
However, for $z/\delta = 0.1$ and $z/\delta = 0.3$, we find that the mean event impulse is significant once within the IBL, at stations $\hat{x}_2$ and $\hat{x}_5$, respectively.
}






\begin{figure}[hbt!]
    \centering
    \includegraphics[width=\linewidth]{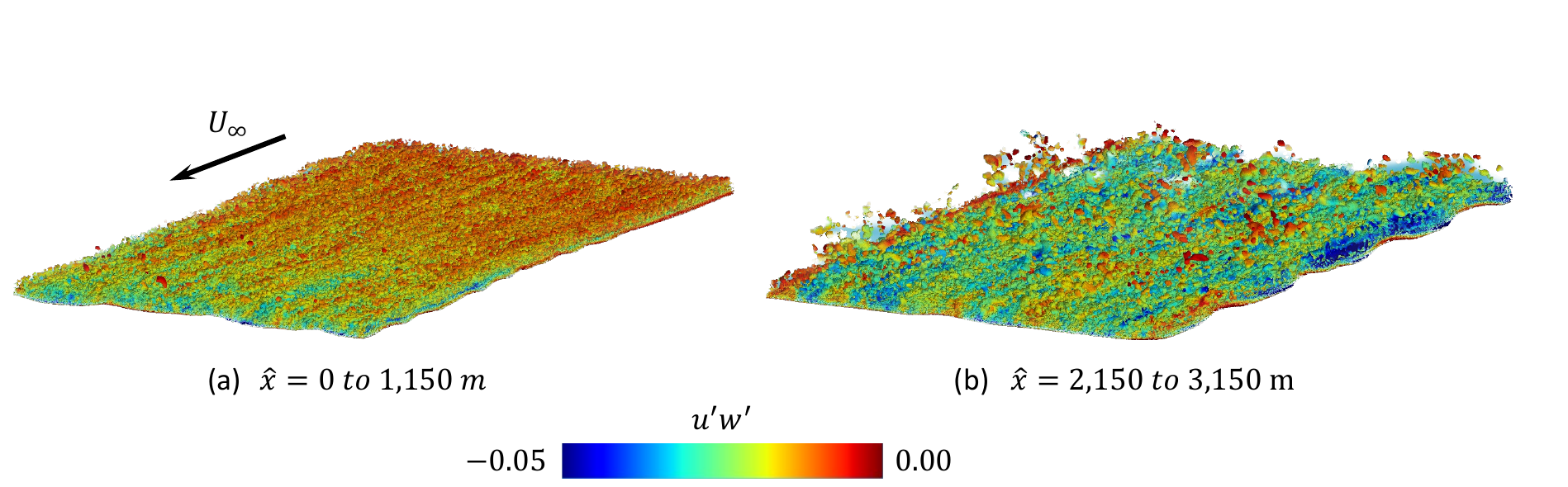}
    \caption{Instantaneous iso-structures of $\lambda_2$-criterion, colored by instantaneous values of $u'w'$. (a) For $\hat{x}$ = 0 - 1,150 m ($\hat{x}_1$ to $\hat{x}_3$), the dunes are only beginning to develop over the first kilometer, resulting in initially smaller structures and lower magnitudes of $u'w'$. (b) Downstream at $\hat{x}$ = 2,150 - 3,150 m ($\hat{x}_6$ and $\hat{x}_7$), the dunes are near their largest values of $\hat{k}_a$. Correspondingly, the flow has adjusted, resulting in large vortex structures and enhanced values of $u'w'$. Flow is from the upper-right to the bottom-left in both (a) and (b).   
    }
    \label{fig:viz}
\end{figure}

Through visualization of the vortex structures in the dune field, we are better able to connect what we have observed with the evolution of event frequency, $T^*_Q$, and $J^*_Q$ after the roughness transition.
We use the $\lambda_2$-criterion proposed by~\citeA{jeong1995on} for visualizing three dimensional coherent vortical  structures, which are colored by instantaneous $u'w'$. 
We compare two $1$-km sections of the dune field: the first $\sim$1 kilometer in which the dunes are initially developing, and approximately between the second and third kilometers, in which the dunes are near their largest values of $\hat{k}_a$. 
Beginning over the first kilometer (Fig.~\ref{fig:viz}a) of the dune field, the flow initially shows low magnitudes of the Reynolds shear stress, with increasing (in magnitude) value as the flow approaches the larger downstream dunes.
Once the flow has reached between the second and third kilometers of the dune field (Fig.~\ref{fig:viz}b), we observe both an increase in the magnitude of the Reynolds shear stress and larger vortex structures away from the wall. 
Here, this increase of $u'w'$ corresponds with larger observed values of $T^*_Q$ and $J^*_Q$ at $z/\delta$ = 0.01 to 0.1.

\subsection{Self-Similarity of $Q_{2}$ and $Q_{4}$ Events in the IBL}
\label{simTurbMotions}

\begin{figure}[hbt!]
    \centering
    \includegraphics[width=\linewidth]{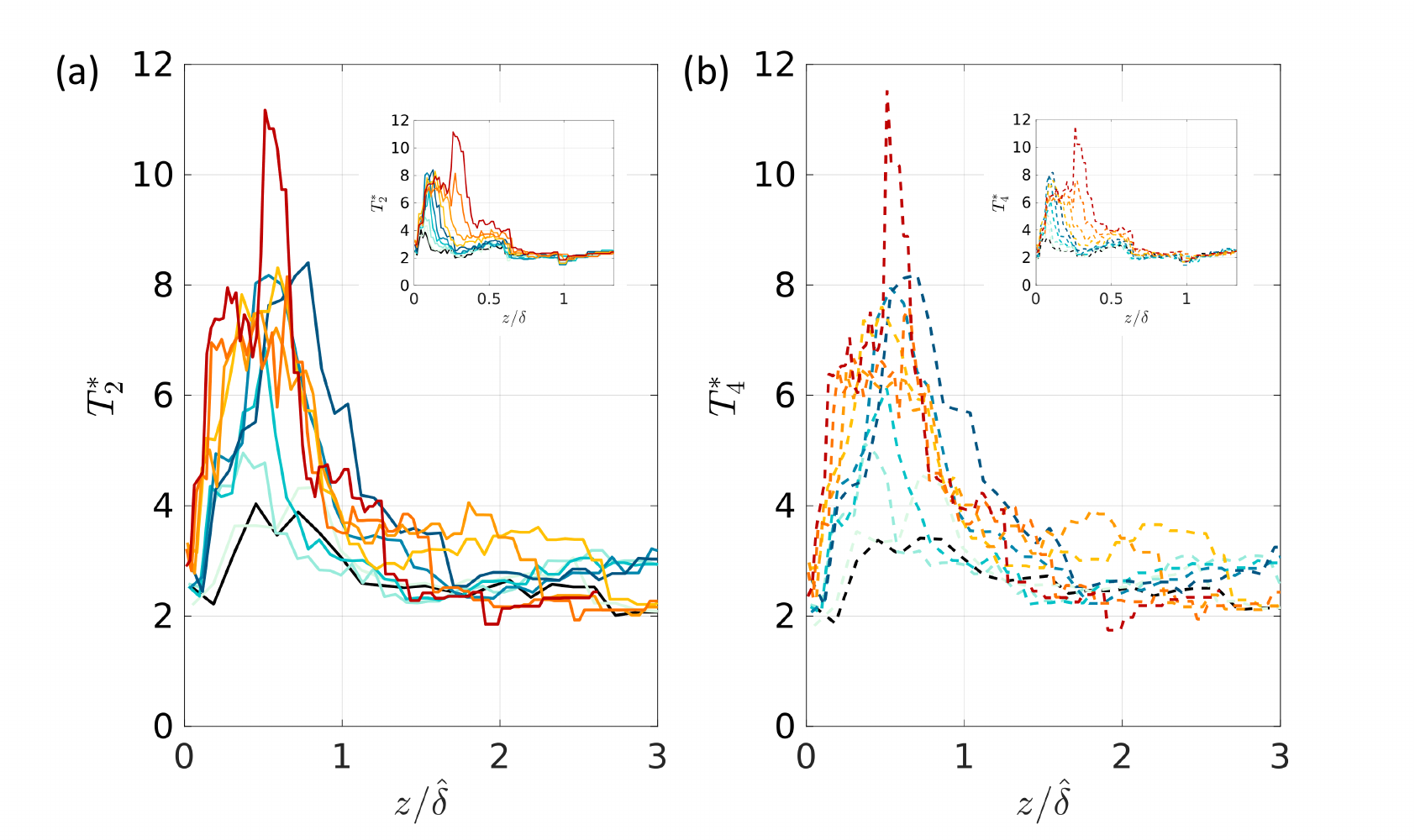}
    \caption{Self-similarity of profiles of $T^*_Q$ for $Q_{2}$ (solid lines) and $Q_{4}$ (dashed lines) after the roughness transition. (a) Profiles of $T^*_{2}$ are plotted with wall-normal location normalized by $\hat{\delta}$, as described previously. Using this normalization results in an alignment of the location where the longest average events occur within the IBL. (b) Profiles of $T^*_{4}$ are plotted in the same manner as (a). Both use the same colors and normalization as Figure~\ref{fig:iblRSS}b. Inset contains the same profiles normalized with a fixed $\delta$, using the same colors and lines styles.}
    \label{fig:TqZ}
\end{figure}

We now observe changes to $T^*_Q$ and $J^*_Q$ with plots of their wall-normal distributions throughout the ABL after the roughness transition, specifically for $Q_2$ and $Q_4$ events.
First, we review $T^*_{2}$ (Fig.~\ref{fig:TqZ}a) and $T^*_{4}$ (Fig.~\ref{fig:TqZ}b), the average time-duration for events associated with ejections and sweeps, respectively.
When plotted against $z/\delta$ (insets of Fig.~\ref{fig:TqZ}a and Fig.~\ref{fig:TqZ}b), there is a clear thickening of the profiles, with event duration increasing and maintaining elevated values at larger $z/\delta$ farther from the roughness transition. 
Magnitudes for $T^*_{2}$ and $T^*_{4}$ are nearly the same at all wall-normal locations, and begin to converge to $T^*_{2} \approx T^*_{4} \approx 2$ shortly above $z/\delta = 0.5$ for all $\hat{x}$.
Noticeably, for both $T^*_{2}$ and $T^*_{4}$, profiles at $\hat{x}_8$ and $\hat{x}_9$ contain an inner peak and a secondary outer peak farther from the wall.
Profiles of $T^*_{2}$ and $T^*_{4}$ are next scaled by the local IBL thickness. We observe a reasonable alignment of $T^*_2$ and $T^*_4$ peak values, which indicates that the largest values of $T^*_{Q}$ occur at a similar scaled elevation. {We determined the mean of the peak locations for $T^*_2$ to be $z/\hat{\delta} = 0.58$, with standard deviation $\sigma = 0.19$. For $T^*_4$, we find a similar mean location of $z/\hat{\delta} = 0.56$, with $\sigma = 0.18$.}
Moreover, the values of $T^*_{2}$ and $T^*_{4}$ begin to decrease and approach their equilibrium state around $z/\hat{\delta} = 1$.

\begin{figure}[hbt!]
    \centering
    \includegraphics[width=\linewidth]{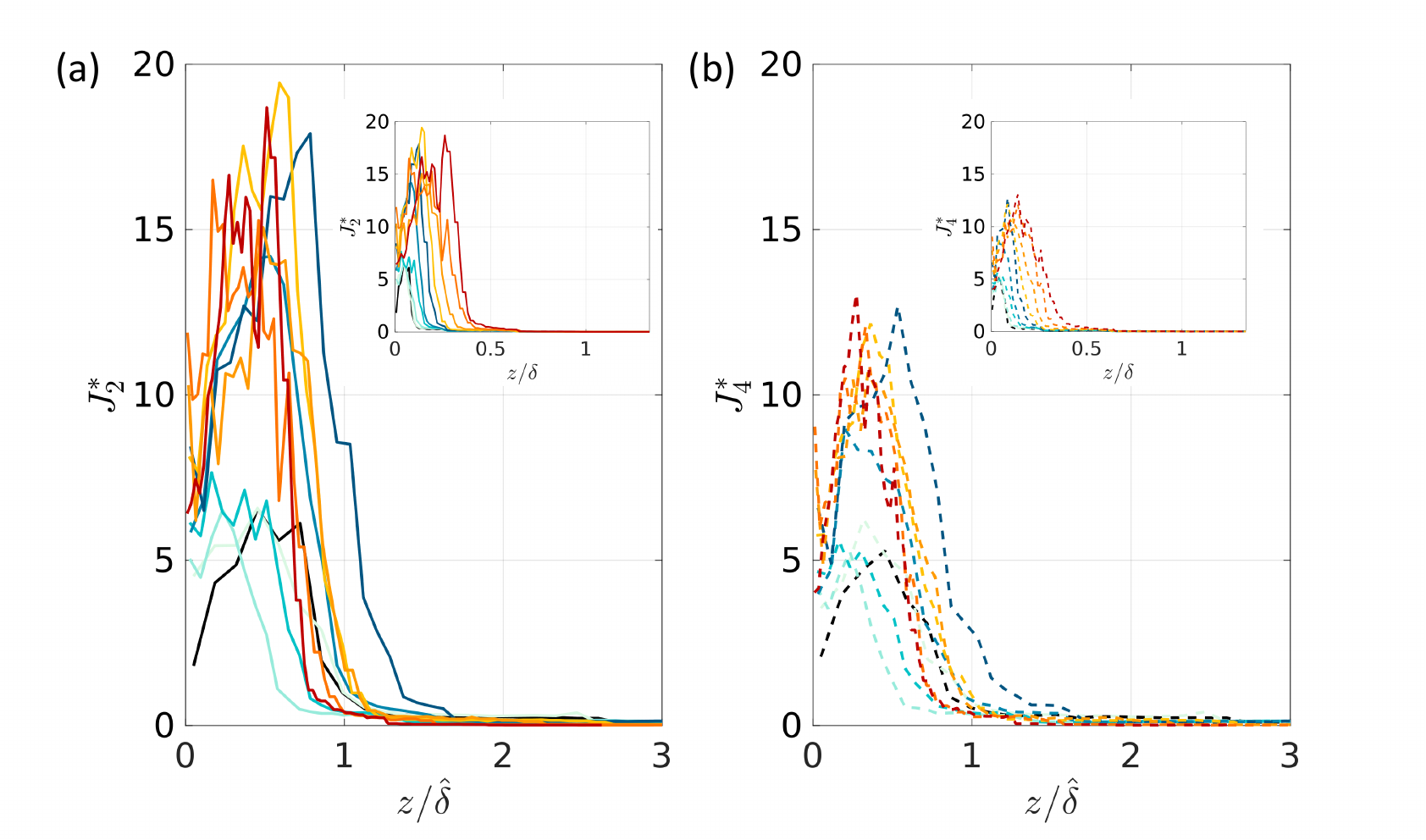}
    \caption{Self-similarity of profiles of $J^*_Q$ for $Q_{2}$ (solid lines) and $Q_{4}$ (dashed lines) after the roughness transition. (a) Profiles of $J^*_2$ are plotted with wall-normal location normalized by $\hat{\delta}$, as described previously. Using this normalization results in an alignment of the location where the largest average impulse for each event occur within the IBL. (b) Profiles of $J^*_{4}$ are plotted in the same manner as (a). Both use the same colors, normalization, and line-styles as Figure~\ref{fig:TqZ}. Inset contains the same profiles normalized with $\delta$, using the same colors and lines styles.}
    \label{fig:JqZ}
\end{figure}

We plot $J^*_{2}$ (Fig.~\ref{fig:JqZ}a) and $J^*_{4}$ (Fig.~\ref{fig:JqZ}b) versus $z/\delta$ (insets of Fig.~\ref{fig:JqZ}).
With increasing distance from the roughness transition, there is an increase in magnitude at similar $z/\delta$ for $J^*_{2}$ and $J^*_{4}$, resulting in a thicker profile further from the wall.
Unlike $T^*_Q$, there are clear differences between the magnitudes of $J^*_{2}$ and $J^*_{4}$. 
For the Alkali Flat and $\hat{x}_1$ through $\hat{x}_3$, values for $J^*_{2}$ and $J^*_{4}$ are similar; however, for $\hat{x}_4$ and further downstream, magnitudes of $J^*_{2}$ become much larger than those for $J^*_{4}$ at similar $z/\delta$.
Additionally, the secondary peak observed for $T^*_{2}$ and $T^*_{4}$ is only observed for $J^*_{2}$.
Now, when $J^*_{2}$ and $J^*_{4}$ are plotted against $z/\hat{\delta}$, as with $T^*_{2}$ and $T^*_{4}$, there is a reasonable self-similarity of the peak location. 
{As with $T^*_2$ and $T^*_4$, we calculate the mean location of the peak values of $J^*_2$ and $J^*_4$, along with the corresponding $\sigma$. 
Here, we find $J^*_2$ has a mean peak location of $z/\hat{\delta} = 0.43$, and $\sigma = 0.22$. 
For $J^*_4$, the mean peak location is $z/\hat{\delta} = 0.32$ and $\sigma = 0.13$.
We note both these locations are lower than their $T^*_Q$ counterparts.}
For both $J^*_{2}$ and $J^*_{4}$, values begin to diminish to zero near $z/\hat{\delta} = 1$.

\section{Discussion}
\label{sec:discussion}

Our main result is the demonstration that the IBL height $\delta_i$ controls the fundamental scales of turbulence within it, even over a natural and heterogeneous roughness transition. 
Here we first consider how our results compare to previous findings in less `messy' systems.
For the smooth-to-rough transition examined here, {our prior work found} that beneath $\delta_i$ there is region of enhanced Reynolds stresses that thickens as the IBL develops downwind~\cite{cooke2024mesoscale}. 
{We determined here that this} increase in energy was accompanied by enhancements of sweep ($Q_{4}$) and ejection ($Q_{2}$) motions (Figs.~\ref{fig:scatter} and~\ref{fig:quadContr}), whose magnitude and wall-normal height change systematically with IBL thickness (Figs.~\ref{fig:TqZ} and~\ref{fig:JqZ}).
\citeA{gul2022experimental} reported {qualitatively} similar observations, for Reynolds stress profiles and sweep and ejection in events, in experiments of IBL development over a step increase in roughness (their Fig. 1f).
We observed that both $Q_{2}$ and $Q_{4}$ events increased in frequency within the IBL, with $Q_{4}$ events occurring at a higher frequency than $Q_{2}$ events; this pattern was also seen in a wind-tunnel model canopy by~\citeA{zhu2007flow}.
As these sweeps originate from large-scale, high-speed fluid motions in the outer-layer, this behavior is to be expected~\cite{bristow2021unsteady,salesky2018buoyancy}. 
We found that the duration and impulse of sweep and ejection events were also much larger within the IBL, with the impulse associated with $Q_{2}$ events being dominant 
(Figs.~\ref{fig:TQ} and~\ref{fig:JQ}). 
{The larger magnitude of $J^*_{2}$ may be the result of increased vertical velocities from the dunes, and enhancement of turbulence from the presence of the IBL.}
In their experiments examining flow over an isolated barchan dune,~\citeA{bristow2021unsteady} found that $Q_{4}$ events dominate closer to the dune while $Q_{2}$ events are dominant farther from the wall.  
Although our results contrast with their findings -- as we do not observe stronger average impulse strength for $Q_{4}$ events nearest wall -- our numerical data do not probe directly above the center line of a single dune. 
More, we do not have access to the flow within the viscous and buffer layers due to the high $Re_\tau$. It is possible also that wall-normal turbulence profiles are different above a train of dunes compared to an isolated dune, because the flow in any location is influenced by the wakes of numerous roughness elements upwind. Examining the difference in flow over an isolated dune, and that same dune embedded in a dune field, would be a useful next step.

The most novel aspect of our study is the quantitative demonstration of how IBL thickness sets the scales of turbulence in the developing flow after a natural smooth-to-rough transition. 
IBL thickness $\delta_i$ sets the eddy turnover time and the energy contained at the largest scales (lowest frequencies in Fig.~\ref{fig:PSD}). 
In other words, the IBL acts as lid that sets the scale of the largest eddies within it, and this scale grows downwind as the IBL thickens.
{We next} examined the turbulence producing motions that contribute to $\langle u'w' \rangle$.
With quadrant analysis (Fig.~\ref{fig:scatter}), we observed the increased magnitude and frequency of $u'$ and $w'$, resulting in a higher frequency of turbulent producing motions further from the wall.
Additionally, we found more frequent $Q_{2}$ and $Q_{4}$ motions with increasing distance from $\hat{x} = 0$ m (Fig.~\ref{fig:freqOfEvents}). 

When using $\delta$ to scale the wall-normal elevation, a thickening of the profiles downstream of the roughness transition was observed, indicating the IBL increased the magnitude of $T^*_Q$ and $J^*_Q$ further from the surface (Inset of Figs.~\ref{fig:TqZ} and~\ref{fig:JqZ}). 
Moreover, when $\hat{\delta}$ was used to scale wall-normal elevation, the point where the `peaks' of $T^*_Q$ and $J^*_Q$ occur collapsed to approximately the same relative point within the IBL (Figs.~\ref{fig:TqZ} and~\ref{fig:JqZ}).
The location of these peaks corresponds to the same location within the IBL, $z/\hat{\delta} \approx 0.5$, where we previously observed a peak negative correlation in the amplitude modulation coefficient ($R_{AM}$)~\cite{cooke2024mesoscale}. 
Previous experimental work has correlated extreme values of sweep and ejection events with increased intermittency~\cite{nakagawa1977prediction,zhang2023multifield}. 
This leads to turbulent motions near the wall that are anti-correlated with the motions of the outer flow~\cite{mathis2009large}, producing a large negative peak in the amplitude modulation coefficient $R_{AM}$.
The convergence of the mean longest and strongest $Q_{2}$ and $Q_{4}$ events, with the peak negative correlation of $R_{AM}$, corroborates these previous findings, and suggests that sweep and ejection events push the intermittency peak away from the wall with increasing $\hat{x}$.

We note that, although we observed an increase in profiles of $T^*_Q$ and $J^*_Q$ (Figs.~\ref{fig:TqZ} and~\ref{fig:JqZ}), {this does not imply that the strength and length of these motions decreases downstream. Further,} when we isolate $z/\delta$ closest to the surface and observe changes in $\hat{x}$ (Figs.~\ref{fig:TQ} and~\ref{fig:JQ}), the values {for $z/\delta = 0.01$ mostly increased downstream, and for $z/\delta = 0.06$, peaked and attained statistical significance} where $\hat{k}_a$ is also largest.
For rough wall-bounded flows, the region of the flow where the effects of roughness are most felt is the roughness sublayer, which extends 2-3$k$ above the roughness and into the flow~\cite{chung2021predicting,jimenez2004turbulent}.
Given the comparatively low height of $\delta_i$ as it develops, the larger magnitudes of $T^*_Q$ and $J^*_Q$ observed at these locations respective to those further downstream could be attributed to the larger values of $\hat{k}_a$. 
Hence, the IBL acts as an intermediate length-scale, bounded below by the roughness and viscous length-scales, and above by the larger length-scales associated with the outer ABL. 

Our simulation results help to explain the observed patterns of dune migration and sediment transport across the White Sands dune field.
There is much evidence for turbulence-producing motions contributing to sediment transport in the near-bed region.
We point out that our simulations do not include particles or particle transport; 
the results presented are used to infer how these motions may affect sediment transport.
{Entrainment of sand by near-surface turbulence extracts momentum from the flow, resulting in an apparent change in the surface roughness parameter $z_0$ when measured in the near-bed region} \cite{field2018controls}. {This effect is confined to the near-surface winds, however, and is not expected to significantly impact the scales of motion resolved in our LES study. 
Because most sand transport conditions are close to the threshold of motion} \cite{jerolmack2010equivalence} {and transport rates are small, the modulation of turbulence by sand entrainment is modest; for this reason, a `frozen surface' approach is common}~\cite{bristow2021unsteady,bristow2022topographic,rana2021entrainment}
Field observations by \citeA{bauer1998event,leenders2005wind,schonfeldt2003turbulence}, and \citeA{sterk1998effect} found sweeping motions were most responsible for initiating and sustaining sediment transport.
For flows over an individual dune, \citeA{wiggs2012turbulent} found evidence that sweeping motions were most responsible for entrainment of sand particles and sediment transport. 
More recently, work from \citeA{tan2023turbulent} investigated sediment transport over the rough surface of the Gobi Desert with a quadrant analysis framework, and determined that, similar to rivers with rough gravel beds, sweep events are major contributors to sediment transport. 
Additionally, experimental wind tunnel work from \citeA{xiao2024role} determined that sufficiently energized sweep events are necessary for entrainment of particles, as weaker sweep events will not prevent the particle from returning to its initial rest position. 
We find that the IBL enhances these sweep and ejection motions. At the start of the dune field the peaks in sweep and ejection events are close to the wall; however, as the IBL thickens, these peaks gradually move away from the wall. 
This pattern may explain our previous observation~\cite{cooke2024mesoscale} that, after an initial increase at the start of the dune field, the boundary stress gradually declines downwind.
A similar pattern was reported for sediment transport by~\citeA{gunn2020macro}.
As the IBL grows, so does the distance between the surface and the location of the {average} longest and strongest events responsible -- and sufficient enough -- for sediment transport{; however, this does not imply that all of the longest and strongest events have completely moved away from the wall}. 
Shortly after $\hat{x}$ = 0 m, the initial IBL growth and subsequent turbulent motions enhance the sediment transport there, but as the IBL thickens these turbulence producing motions migrate away from the wall, resulting in the gradual decline of sediment transport downwind. 
These turbulent motions also influence the transport of other particulates, including dust and aerosols. 
In the saltation process, the impact of individual sand grains on the bed can release dust trapped within~\cite{rana2021entrainment,shao2020dependency,klamt2024saltation}.
In their study, \citeA{rana2021entrainment} showed that regions of high momentum -- generally characterized by sweeping motions -- enable both saltation and, indirectly, dust entrainment. 
Additionally,~\citeA{shao2020dependency} confirmed the dependence of dust particle size distribution on $u_\tau$, and indicated that stronger turbulence, which results in larger mean values of $u_\tau$ and greater variance, would result in increased saltation-bombardment and dust emission.
As our results indicate, the increased frequency of sweeping motions has the potential to increase dust emission, especially nearest the roughness transition.
Moreover, as the strongest motions move further from the wall, so too could the potential for the entrained dust to be carried away from the surface by these motions. 
Although all the findings here pertain to simulations of turbulent flow over a natural dune field, the discovery that IBL thickness scales the profile of turbulent producing motions is a result that we expect may generalize to other ABL flows over roughness transitions. 
For urban topography where low-lying buildings transition to high-rise skyscrapers, this rough-to-rougher transition will see an enhancement to turbulence producing motions and changes to thermal stratification, with the capacity to augment the transport of momentum, particles, heat, and moisture \cite{sessa2020thermal,rios2023turbulence}. 
The developing IBL at the land-ocean interface exerts a control on moisture transport \cite{jiang2021characteristics}. 
Finally, roughness transitions such as field to forest canopy should also produce similar behaviors, influencing the transport of moisture and CO$_2$, as turbulent mixing over forest canopies is enhanced \cite{baldocchi2003assessing}.
The deployment of eddy flux towers should explicitly account for IBL development, which produces spatially varying turbulence motions over distances of many kilometers. 

\section{Conclusion}
\label{sec:Conclusion}

We performed Wall-Modeled Large-Eddy Simulation (WMLES) of a neutrally buoyant Atmospheric Boundary Layer (ABL) encountering a roughness transition between a smooth playa and a spatially heterogeneous dune field.
Our simulations captured the development of an Internal Boundary Layer (IBL) which forms at the inception of the dune field. 
Using observations of the energy frequency spectrum of the streamwise velocity fluctuations $E(\omega)$ at multiple locations downstream of the roughness transition, we show how the IBL sets the low frequencies (large scales) of turbulence.
Additionally, we calculate a frequency associated with the IBL, and find it correlates well with the scaling break in the energy frequency spectrum, typically associated with the largest-scale eddy turnover time. 
As a result of the IBL setting the largest scales, we show how the IBL enhances turbulence producing motions throughout, especially ejection ($Q_2$) and sweep ($Q_4$) events.
Moreover, these events display a self-similarity at subsequent downstream locations, in both the average event time-duration ($T^*_Q$) and impulse strength ($J^*_Q$). 
As the location of the longest and strongest of these events migrates away from the wall with the growing IBL, the enhancement of sediment flux, and transport of other materials, over the initial portion of the dune field is lost downstream. 
For ABL flows encountering roughness transitions, $\delta_i$ is clearly a prominent mesoscopic length scale; its spatial growth scales the profile of many turbulent characteristics that are not captured with inner- or outer-scalings. 
It would be beneficial to deploy this scaling for other roughness transitions -- rough-to-smooth, rough-to-rougher, and vice versa -- to see if it is universal.
Additionally, our WMLES does not capture the near-surface flow characteristics, as the high $Re_\tau$ incurs a high computational cost. A direct numerical simulation, or experimental study, at a more moderate $Re_\tau$ would allow for examining the efficacy of this scaling closer to the wall.

\section{Open Research}
The authors have made available the data and scripts required to recreate Figures and conduct the analysis in the present article. The data and scripts for Figure 6 can be found in the GitHub Repository \url{https://github.com/JCooke188/Dune_AM}, and the data and scripts for Figures 7--12, 14, and 15 can be found in the Github Repository \url{https://github.com/JCooke188/Dune_QA}. The DOI for each repository is \url{doi.org/10.5281/zenodo.10476149}~\cite{https://doi.org/10.5281/zenodo.10476149} and \url{doi.org/10.5281/zenodo.11521385}~\cite{https://doi.org/10.5281/zenodo.11521385}, respectively.

\acknowledgments
G.P. and J.C. acknowledge the support from the University of Pennsylvania (faculty startup grant and the Fontaine Fellowship) and the National GEM Consortium Fellowship.
D.J.J. was supported by NASA PSTAR (Award 80NSSC22K1313).
We would also like to acknowledge Prof. Andrew Gunn for helpful discussions related to his work at White Sands, and for providing his experimental data. 
The authors declare no conflict of interest related to this work, financial or otherwise.


%
%



\bibliography{dunesBib}

%
%
%
%
%

\end{document}